\documentclass[fleqn,usenatbib]{mnras}

\usepackage{newtxtext,newtxmath}
\usepackage[T1]{fontenc}
\usepackage{tabularx}
\usepackage[normalem]{ulem}
\usepackage{multirow}
\usepackage{makecell}
\usepackage{textcomp}
\usepackage{subcaption}
\usepackage{dcolumn}
\usepackage{multirow}
\usepackage{ulem}
\usepackage{color}
\usepackage{hyperref}
\usepackage{url}
\usepackage{enumitem}

\DeclareRobustCommand{\VAN}[3]{#2}
\let\VANthebibliography\thebibliography
\def\thebibliography{\DeclareRobustCommand{\VAN}[3]{##3}\VANthebibliography}

\usepackage{graphicx}	
\usepackage{amsmath}	
\hypersetup{
    colorlinks=true,
    citecolor=blue,
    urlcolor=blue,
} 

\title[Probing nuclear physics with supernova gravitational waves and machine learning]{Probing nuclear physics with supernova gravitational waves and machine learning}

\newcommand{\be}{\begin{equation}}
\newcommand{\ee}{\end{equation}}
\newcommand{\bea}{\begin{eqnarray}}
\newcommand{\eea}{\end{eqnarray}}

\definecolor{mediumpurple}{rgb}{0.58, 0.44, 0.86}

\newcommand{\gfirstscore}{$0.78\pm0.08$}
\newcommand{\gfirstFTscore}{$0.73\pm 0.05$}

\newcommand{\gsecscore}{$0.85\pm0.04$}
\newcommand{\gsecFTscore}{$0.82\pm 0.05$}

\newcommand{\gthirdscore}{$0.87\pm0.03$}
\newcommand{\gthirdFTscore}{$0.84\pm0.04$}

\newcommand{\gthirdbscore}{$0.80\pm0.06$}
\newcommand{\gthirdbFTscore}{$0.79\pm0.04$}

\author[A. Mitra et al.]{
A. Mitra$^{1,2,3,4}$,
D. Orel$^{5}$,
Y. S. Abylkairov$^{6}$\thanks{E-mail: sultan.abylkairov@nu.edu.kz},
B. Shukirgaliyev$^{6,7,8,9}$,
and E. Abdikamalov$^{2,6}$
\\
$^{1}$Center for Astrophysical Surveys, National Center for Supercomputing Applications, University of Illinois Urbana-Champaign, Urbana, IL, 61801, USA\\
$^{2}$Department of Physics, Nazarbayev University, 53 Kabanbay Batyr ave, 010000 Astana, Kazakhstan\\
$^{3}$Department of Astronomy, University of Illinois at Urbana-Champaign, Urbana, IL 61801, USA\\
$^{4}$Kazakh-British Technical University, 59 Tole Bi Street, 050000 Almaty, Kazakhstan\\
$^{5}$Department of Computer Science, Nazarbayev University, 53 Kabanbay Batyr ave, 010000 Astana, Kazakhstan\\
$^{6}$Energetic Cosmos Laboratory, Nazarbayev University, 53 Kabanbay Batyr ave, 010000 Astana, Kazakhstan\\
$^{7}$Heriot-Watt International Faculty, Zhubanov University, 263 Zhubanov brothers str, 030000 Aktobe, Kazakhstan\\
$^{8}$Fesenkov Astrophysical Institute, 23 Observatory str, 050020 Almaty, Kazakhstan\\
$^{9}$Department of Computation and Data Science, Astana IT University, 55/11 Mangilik El ave, 010000 Astana, Kazakhstan
}

\date{Accepted XXX. Received YYY; in original form ZZZ}

\pubyear{2023}

\begin{document}
\label{firstpage}
\pagerange{\pageref{firstpage}--\pageref{lastpage}}
\maketitle

\begin{abstract}
Core-collapse supernovae are sources of powerful gravitational waves (GWs). We assess the possibility of extracting information about the equation of state (EOS) of high density matter from the GW signal. We use the bounce and early post-bounce signals of rapidly rotating supernovae. A large set of GW signals is generated using general relativistic hydrodynamics simulations for various EOS models. The uncertainty in the electron capture rate is parametrized by generating signals for six different models. To classify EOSs based on the GW data, we train a convolutional neural network (CNN) model. Even with the uncertainty in the electron capture rates, we find that the CNN models can classify the EOSs with an average accuracy of about 87 percent for a set of four distinct EOS models.
\end{abstract}


%%%%%%%%%%%%%%%%%%%%%%%%%%%%%%%%%%%%%%%%%%%%%%%%%%

\section{Introduction}

Core-collapse supernovae are the powerful explosions that take place at the end of lives of massive stars. A fraction of the gravitational binding energy released in stellar core-collapse is transferred to the ejection of the stellar envelope. Despite decades of effort, the exact details of how this happens remain unknown \citep[e.g.,][for recent reviews]{janka:16a, mueller:20review, Burrows21review}. The supernovae produce powerful bursts of photons, neutrinos and gravitational waves (GWs) \citep[e.g.,][]{Nakamura:2016kkl}. The future multi-messenger observations of CCSNe will provide unprecedented insight into these phenomena \cite[e.g.,][]{Warren20}. 

As massive stars evolve, they go through all stages of nuclear burning, synthesizing heavier and heavier elements. At the end of the process, iron core forms \citep{whw:02}. The core is supported by the pressure of degenerate electrons. Upon reaching their effective Chandrasekhar mass, the iron core becomes unstable and starts collapsing. When nuclear densities are reached, the strong nuclear force abruptly halts the collapse. The inner core rebounds and collides with still-infalling outer parts, launching a shock wave. The dissociation of heavy nuclei and neutrino cooling quickly drains the kinetic energy of shock, stalling the shock at $\sim \! 150 \, \mathrm{km}$. To produce a supernova explosion and leave behind a stable protoneutron star (PNS), the shock must revive and expel the stellar envelope within a second \cite[e.g.,][]{ertl:16, daSilvaSchneider20}.  

The PNS cools and contracts via the emission of neutrinos. Most of these neutrinos escape to infinity, but a small fraction is absorbed behind the shock. The deposited energy heats the post-shock medium, pushing the shock outwards \cite[e.g.,][]{janka:01}. Moreover, the neutrino heating drives convection in that region \citep{herant:94, bhf:95, janka:96}. In addition, the shock front becomes unstable to large-scale oscillations known as standing accretion shock instability (SASI) \citep{blondin:03, foglizzo:06, iwakami:09, scheck:08, fernandez:09a, mueller:12b}. These effects push the shock forward \citep{murphy:13, radice:16a, fernandez:15a}. The expanded post-shock flow absorbs more neutrinos, leading to more shock expansion \citep[e.g.,][]{mueller:15}. This paradigm is called the neutrino mechanism for CCSN explosions. 

The vast majority of stars are found to be slow rotators \citep[e.g.,][]{heger:05, mosser:12, popov:12, deheuvels:14}. These stars are expected to explode via the neutrino mechanism and the rotation is unlikely to have a significant impact on the explosion dynamics. However, a small fraction of massive stars may possess rapid rotation \citep{fryer:05, woosley:06, yoon:06, demink:13}. In these stars, the PNSs are born with immense $\lesssim 10^{52} \, $ erg rotational kinetic energy. Magnetic fields transfer a part of this energy to the shock front via the so-called magneto-rotational mechanism \citep{burrows:07b, winteler12, moesta:14b, obergaulinger:20, kuroda:20}. This mechanism is thought to be responsible for the extremely energetic hypernova explosions \citep[e.g.,][]{woosley_bloom:06}. Note that magnetic fields may play a significant role in slowly or non-rotating models too \citep[e.g.,][]{endeve:12, mueller:20b, varma23}. 

In addition, if a quark deconfinement phase transition takes place inside the PNS, the PNS may undergo a ``mini collapse", launching a second shock wave. This helps the first shock to expel the stellar envelope \citep{sagert:09, zha:21}. However, whether this happens is unclear as it relies on uncertain assumptions about the properties of high-density matter.

One of the promising ways of learning more about CCNSe is by detecting gravitational waves (GWs) from these events. While GWs from mergers of black holes and neutron stars are now observed routinely, we are waiting for the first detection of GW from CCSNe \citep{abbott:20ccsn, Lopez21, Antelis22, szczepanczyk2023optically}. The GWs coming from Galactic CCSNe should be detectable with the current observatories \citep[e.g.,][]{gossan:16}. CCSNe are expected to occur once or twice per century in our galaxy \citep[e.g.,][]{adams2013observing}. The future detectors will be sensitive to observe events from longer distances \citep[e.g.,][]{Srivastava:2019fcb}, which should increase the detection event rate of CCSNe. 

GWs are generated by asymmetric dynamics that take place in the inner regions \citep{mueller:13, kuroda:17, radice:19gw, andresen:17, mueller:97, powell:20, mezzacappa23b, Vartanyan23}. Since the GWs carry information about the sources, one can extract parameters of the source from the signal \citep{Szczepanczyk21, Powell22, bruel2023inference, pastor2023bayesian, Afle23, Yuan23reconstruction, Mori23}. In particular, we may be able to probe the explosion mechanism \citep{Logue12, Powell16, kuroda:17, Chan20, Saiz-Perez22}, measure core rotation \citep{abdikamalov:14, Engels14, Edwards14, hayama16, Afle21}, and probe the structure of the PNS and the parameters of high-density nuclear matter \citep{richers:17, morozova:18, Torres-Forne19, pajkos21, Sotani21, Andersen21, Wolfe23, casallas2023characterizing}. Note that constraints on nuclear EOS can also be obtained using GW signals from NS mergers \cite[e.g.,][]{baiotti2017binary, radice2018gw170817, carson2019future, bauswein2020equation, pacilio2022ranking, puecher2023unraveling, iacovelli2023nuclear}. See, e.g., \citet{lattimer2023constraints} for a recent review on nuclear EOS constraints from experiments and neutron star observations. 

In this work, we explore if it is possible to extract any information about the EOS of high-density nuclear matter from the GW signal. As we explain below, we focus on the so-called bounce signal from rotating models. The dynamics, and thus the GW signal, depends on the EOS. Using numerical simulations, we generate a large number of GW signals using different EOS models. For each EOS, we produce up to hundreds of signals that correspond to different rotational configurations and electron capture rates during collapse. We train machine learning (ML) models to classify these signals based on their EOSs. We then estimate how well the ML model can infer the EOS information from a GW signal alone. The aim of this exercise is to answer the question: when the real CCSN GW signal will be detected, will such models be able to tell which EOS model closely represents the detected signal? If the answer is positive, this would mean that ML models will be able to put constraints on the properties of high-density nuclear matter.

The reason we focus on the bounce signal from rapidly rotating models is the following. In these models, the centrifugal force causes strong non-radial deformation of the bouncing core. The PNS is then born with strong perturbation. This drives ring-down oscillations in the post-bounce phase \citep{ott:12}. These oscillations decay within $\sim 10$ ms due to hydrodynamic damping \citep{fuller:15}. The bouncing core generates a spike in GR strain (aka bounce GW signal), while the post-bounce PNS oscillations generate GWs at the frequency of these pulsations \cite[e.g.,][]{abdikamalov:22review}. This signal can be modeled relatively easily with general relativistic simulations using a simple deleptonization scheme \citep{liebendoerfer:05b}. This allows us to create a large set of GW waveforms with moderate computational cost. 

Previously, \cite{Edwards21} and \cite{Chao22} performed machine learning classification of a large set of GW signals that correspond to 18 different EOSs generated by \citet{richers:17}. We extend these studies by further analyzing the impact of the uncertainty in the electron capture rate during collapse. To parametrize the uncertainty, consider six different electron capture models and generate additional waveforms for a subset of four EOSs that produce distinct signals from each other. Even with the uncertainty in the electron capture rates, we find that the machine learning model can classify these EOS with an average accuracy of about 87 percent.

This paper is organized as follows. In Section \ref{sec:method} we describe our methodology. In Section \ref{sec:results}, we present our results. Finally, in Section \ref{sec:conclusion}, we summarize of results and provide conclusions. 

%%%%%%%%%%%%%%%%%%%%%%%%%%%%%%%%%%%%%%%%%%%%%%%%%%%%%%%%%%%%

\section{Methodology}
\label{sec:method}

We use supervised deep learning technique to perform EOS classification based on GW data. Our pipeline is comprised of 1-dimensional convolution neural network (CNN) algorithm  \citep{lecun2015deep, lecun1989backpropagation, lecun1998gradient, krizhevsky2012imagenet, simonyan2014very, szegedy2015going, he2016deep, huang2017densely}. Following \citet{Edwards21}, it is composed of nine layers, with 3 convolution layers, 3 maxpool layers, 3 dense layers, and with an in-between flattening layer. We use ReLU and softmax activation functions. The model parameters are summarized in Table~\ref{tab:NNmodel}. We utilize the {\tt TensorFlow} framework \citep{abadi2016tensorflow} and {\tt scikit learn} library \citep{pedregosa2018scikitlearn}. In Table~\ref{tab:hyperp} we summarize the hyperparameters used for training.

\begin{table}
\centering
\begin{tabular}{lll}
\hline
Layer (type)                & Output Shape             & Activation  \\
\hline
Convolution 1D              & (None, 1178, 32)         & ReLU     \\
Max Pooling 1D & (None, 589, 32)          &         \\
Convolution 1D            & (None, 587, 64)          & ReLU     \\
Max Pooling 1D & (None, 293, 64)          &         \\
Convolution 1D            & (None, 291, 128)         & ReLU    \\
Max Pooling 1D & (None, 145, 128)         &         \\
Flatten            & (None, 18560)            &         \\
Dense                & (None, 512)              & ReLU  \\
Dense              & (None, 256)              & ReLU   \\
Dense              & (None, 18)               & Softmax     \\
\hline
\hline
\end{tabular}
\caption{Parameters of the 1D CNN model architecture used in this work.}
\label{tab:NNmodel}
\end{table}

\begin{table}
\centering
\begin{tabular}{ll}
\hline
\textbf{Hyperparameters} & \textbf{Value} \\
\hline
Batch Size & 32 \\
Loss Function & Sparse Categorical Crossentropy\\
Optimizer & Adam \\
Activation function & ReLU, softmax \\
Learning Rate & 0.001 \\
Number of Epochs & 20 \\
Evaluation & Accuracy and F1  score \\
\hline
\hline
\end{tabular}
\caption{Summary of the hyperparameters of our CNN model.}
\label{tab:hyperp}
\end{table}

We use sparse categorical cross entropy loss function and Adaptive Moment Estimation (Adam) optimizer for training \citep{sonderby2016categorical,goodfellow2016deep,kingma2014adam}. To enable learning of complex patterns, we use the Rectified Linear Unit (ReLU) function \citep{hanin2019universal, pinkus1999approximation}.

%\sout{Sparse categorical cross-entropy is a classification loss function commonly used in machine learning for tasks where target labels are represented as integers, instead of one-hot encoded vectors. It quantifies the difference between predicted and true probabilities . Adam is an optimization algorithm, which combines techniques from both momentum and RMSprop to adaptively adjust learning rates for each parameter. These features make Adam efficient and versatile for training neural networks (NNs)} \citep{goodfellow2016deep}.

%\sout{The Softmax function transforms a vector of $K$ real values into another vector of $K$ real values, ensuring that the resulting values sum to $1$. This is typically applied in the final layer of an NN-based classifier and it represents the probability distribution for a set of potential outcomes. }\citep{cybenko1989approximation, funahashi1989approximate, hornik1989multilayer}.  

We use the accuracy and F1 score as the evaluation metric. The accuracy metric calculates the ratio of the correct predictions to the total predictions made:
\begin{equation}
\text{Accuracy} = \frac{\text{Number of Correct Predictions}}{\text{Total Number of Predictions}}.
\end{equation}
This metric is useful in scenarios with balanced classes and roughly equal costs for false positives and false negatives \citep{2020arXiv201204193C}.

The F1 score is a  harmonic mean of precision and recall~\citep{1361981469291709056}. It is calculated by taking twice the product of precision and recall and dividing it by their sum:
\begin{equation}
    F1 = 2 \times \frac{\text{Precision} \times \text{Recall}}{\text{Precision} + \text{Recall}},
\end{equation}
where precision and recall are defined as 
\begin{equation}
    \text{Precision} = \frac{\text{True Positives}}{\text{True Positives} + \text{False Positives}},
\end{equation}

\begin{equation}
    \text{Recall} = \frac{\text{True Positives}}{\text{True Positives} + \text{False Negatives}}
\end{equation}

\begin{figure}
    \centering
    \includegraphics[scale=0.48]{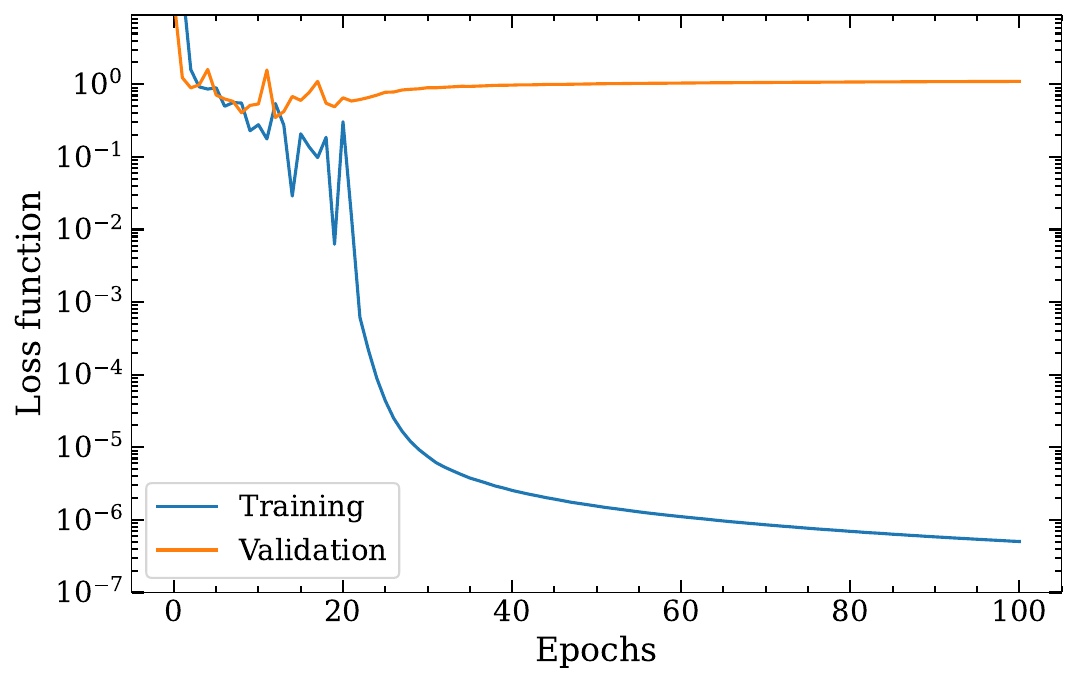}
    \caption{Loss function as a function of the epochs for group 1 dataset. In our analysis, we use models trained at epoch $20$.} 
    \label{fig:loss}
\end{figure}

Figure~\ref{fig:loss} shows the evolution of the loss functions for the training and validation over epochs for the group 1 dataset (described later in this section). These functions experience rapid decrease during the first $\sim 3$ epochs. After that, we see mild fluctuations until epoch $\approx 20$. During this period, the training loss decreases by $\sim 1$, while the validation loss does not change much (except experiencing fluctuations). After epoch $20$, the validation loss increases gradually from $\approx 0.65$ at epoch $20$ to $\approx 1.1$ at epoch 100, while the training loss decreases from $\approx 0.3$ to $\approx 5 \times 10^{-7}$. This is a sign of overfitting. A similar overfitting was also observed in \cite{Edwards21}. For this reason, we use $20$ epochs for training our models in the rest of the paper. 

We obtain the GWs from general relativistic hydrodynamics simulations using the {\tt CoCoNuT} code \citep{dimmelmeier:05MdM}. We model neutrino effects using the $Y_e(\rho)$ parametrization \citep{liebendoerfer:05b} in the collapse and bounce phase, after which we switch to a leakage scheme. The $Y_e(\rho)$ parametrization assumes that the electron fraction during collapse phase depends only on density \citep{mueller:09phd}. Since the stellar core is expected to remain rotationally symmetric during collapse and early post-bounce phase \citep{ott:07cqg}, the simulations are performed in axial symmetry. We do not include magnetic fields as they have little impact on the dynamics of the core during collapse and early post-bounce phase \cite[e.g.,][]{obergaulinger:06}. 

We consider two sets of models. In the first set, we take the GW data from simulations of \cite{richers:17} for 18 different EOSs. A summary of the EOS parameters is provided in Table 1 of \cite{richers:17}. Rotation is imposed on the initial stellar core according to 
\begin{equation}
    \label{eq:rot_law}
    \Omega(\varpi) = \Omega_0 \left[ 1 + \left(\frac{\varpi}{A}\right)^2\right]^{-1},
\end{equation}
where $\varpi$ is the cylindrical radius, $\Omega_0$ is the central angular velocity, and $A$ is a degree of differential rotation. By varying the latter two parameters, we obtain up to $98$ different rotational configurations, ranging from slow to fast rotation, for each EOS and $Y_e (\rho)$ model. Some of the models with extremely rapid rotation do not collapse due to the excessive centrifugal force. The list of these models is given in Table 3 of \cite{richers:17}. These models do not emit significant bounce GW signals, so we exclude them from our analysis. A similar approach was adopted by \cite{Chao22}. This group contains 1704 waveforms in total. We refer to this dataset as group 0 hereafter. 

Note that for a given angular momentum distribution with respect to the enclosed mas coordinate in the core, different progenitor stars produce similar bounce GW signals \citep{ott:12,  mitra23}. For this reason, we focus on one progenitor star model s12 of \cite{woosley:07}. 

In the second set, we take four of the 18 EOSs from \cite{richers:17} and perform simulations using additional electron fraction $Y_e(\rho)$ profiles. This is done to parametrize the uncertainty in the electron capture rate, which affects the values of $Y_e$ and thus the dynamics of stellar collapse \citep[e.g.,][]{Hix:03}. See \citet{langanke2021electron} for a recent review of electron capture rates in supernovae. For each combination of EOS and $Y_e(\rho)$ profile, we obtain 80 different rotational configurations by excluding those out of 98 model that do not collapse. 

The four EOSs that we consider are {\tt LS220} \citep{lseos:91},  {\tt GShenFSU2.1} \citep{gshen:11b}, {\tt HSDD2} \citep{hempel:10, hempel:12}, and {\tt SFHo} \citep{steiner:13b}. Based on experimental and neutron star mass measurement constraints, these four EOSs represent relatively realistic EOSs from the set of \cite{richers:17}, as can be seen in Fig. 1 of \cite{richers:17}. At the same time, these four EOSs produce relatively distinct peak GW signal frequencies, as can be seen in Fig. 10 of \cite{richers:17}. 

\begin{figure}
    \centering
    \includegraphics[scale=0.47]{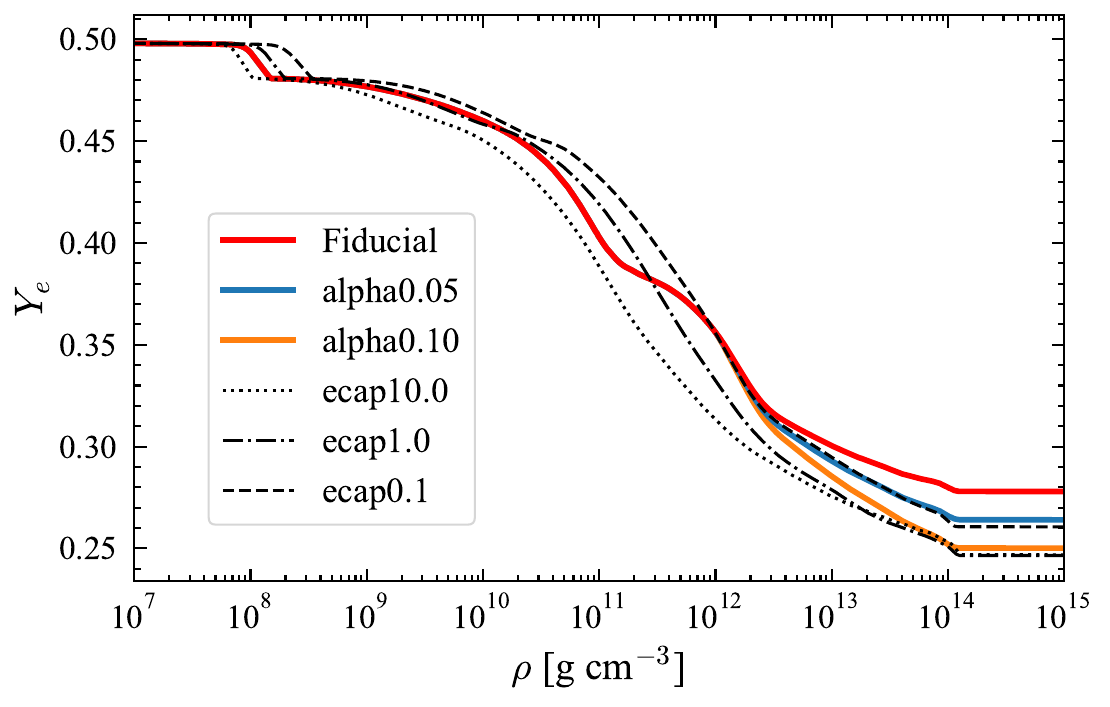}
    \caption{Electron fraction profiles for the \texttt{SFHo} EOS. The red curve represents the fiducial $Y_e(\rho)$ profile, while the blue and orange curves represent the profiles adjusted using the formula (\ref{eq:dye}) for $\alpha = 0.05$ and $0.1$, respectively. The dashed, dashed-dotted, and dotted lines represent the $Y_e(\rho)$ curves obtained from \texttt{GR1D} simulations using the electron capture rates of  \citet{Sullivan16}, scaled by a factor of $0.1$, $1$, and $10$, respectively.}

    \label{fig:Ye_profiles}
\end{figure}

We consider three groups of $Y_e(\rho)$ profiles. In group 1, which consists of 320 waveforms, we take the $Y_e(\rho)$ profiles from  \cite{richers:17}. We refer to these as fiducial profiles. In group 2, we add two $Y_e(\rho)$ profiles, which are obtained by adjusting the fudicial $Y_e(\rho)$ profiles above density $\rho_1 = 10^{12}  \, \mathrm{g/cm^3}$ by a factor of 
\begin{equation}
\label{eq:dye}
\delta Y_e (\rho) = \bigg(1 - \alpha \frac{\log \rho - \log \rho_1}{\log \rho_2 - \log \rho_1} \bigg) \ Y_e (\rho), \quad \rho_1 < \rho < \rho_2
\end{equation}
where $\rho_2 = 10^{14} \, \mathrm{g/cm^3} $. This relation is motivated by the fitting formula (1) of \citet{liebendoerfer:05b}. We consider two different values of $\alpha$ of $0.05$ and $0.1$. This means that the $Y_e$ value in the stellar core will be $5\%$ and $10\%$ smaller than the fiducial model. In total, group 2 has $960$ waveforms. Finally, for group 3, we add three more $Y_e(\rho)$ profiles obtained from {\tt GR1D} simulations \citep{oconnor:15a} using the electron capture rates of \cite{Sullivan16}, scaled by a factor of $0.1$, $1$, and $10$, as was done by \cite{richers:17} for the {\tt SFHo} EOS. The plot of corresponding values of $Y_e$ as a function of $\rho$ is shown in Fig.~\ref{fig:Ye_profiles}. In total, we have 1200 waveforms in group 3. Note that group 1 is a subset of group 2, which, in turn, is a subset of group 3. 

For an additional test, we create group 3b by randomly removing $240$ waveforms corresponding to the {\tt SFHo} EOS from group 3, resulting in a total of $960$ waveforms, the same number as in group 2. This will help us assess how the number of waveforms and variations of $Y_e(\rho)$ affect the classification accuracy.

For each of the groups, we randomly shuffle the waveforms such that $80\%$ are used as a training set and $20\%$ as a test set. At the same time, we ensure a balanced representation of classes in both the training and test sets. In the case of group 3, where there are more \texttt{SFHo} samples than others, we proportionally increase the number of these EOS in both training and test sets to align with the general distribution. This procedure is repeated 10 times, and the EOS classification results that we report below are averaged over these 10 realizations. The error measurements are expressed in terms of the corresponding standard deviation.

Each GW signal is labeled with a corresponding EOS that we aim to classify. To quantify the result, we use the accuracy metric defined as the fraction of correct EOS classifications. The results presented below are computed using the time series data in real space. A complementary analysis in the Fourier space is provided in Appendix \ref{sec:fourier}.

Before ML analysis, we follow \citet{Edwards21} and apply a Tukey window with $\alpha=0.1$ and Butterworth filter with order $10$ and attenuation $0.25$ on all data \citep{blackman1958measurement, SmithGossett1984}. We adjust the time axis so that $t=0$ ms corresponds to the time of bounce. The latter is defined as the time when then entropy along the equator exceeds $3 k_b \, \mathrm{baryon}^{-1}$, which is the result of the heating by the shock formed at bounce. Our GW data is sampled at the rate of $0.01$ ms. We performed comparison with sampling rates of $0.2$ and $0.1$ ms and find no statistically significant differences between these sampling rates, as we show in Appendix \ref{sec:sampling}. This is an expected outcome as most of the physical (and EOS-dependent) signal is contained below $\sim 1$ kHz \citep[e.g.,][]{dimmelmeier:08}.

%%%%%%%%%%%%%%%%%%%%%%%%%%%%%%%%%%%%%%%%%%%%%%%%

\section{Results}
\label{sec:results}

\begin{figure}
    \centering
    \includegraphics[scale=0.48]{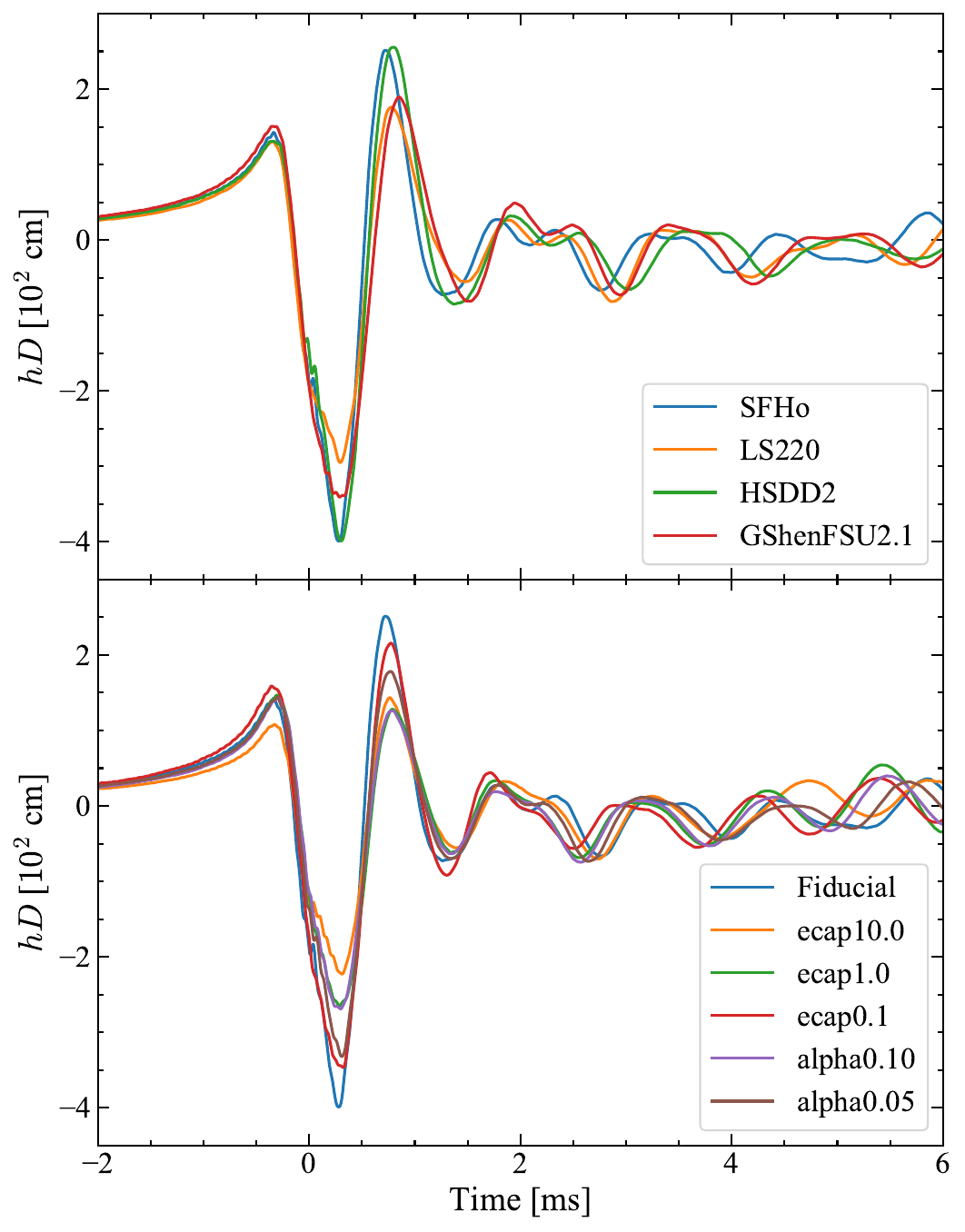}
    \caption{GW signals with the same rotation profile $A=634$ km, $\Omega_0=5.0$ rad s$^{-1}$ and with $T/|W| \approx 0.06$. The top panel displays the GW signal for four EOSs with a fiducial $Y_e(\rho)$ profile, while the bottom panel shows the GW signal for the {\tt SFHo} EOS with various $Y_e(\rho)$ profiles shown in Fig.~\ref{fig:Ye_profiles}.  
    }
    \label{fig:Waveforms}
\end{figure}

We first look at the main qualitative features of the bounce and ring-down GW signal. For a given progenitor model and rotational configuration, the dynamics depends on the EOS and the electron fraction profile. Approximately, the bounce GW amplitude can be estimated as \citep{richers:17}
\begin{equation}
\label{eq:hestimate}
    h D \sim \frac{G^2}{c^4}  \frac{M^2}{R}  \frac{T}{|W|},
\end{equation}
where $D$ is the distance to the source. The dependence on the EOS and $Y_e(\rho)$ profile enters this equation via the ratio $M^2/R$, where $M$ and $R$ are the mass and radius of the inner core at bounce. The inner core mass scales as $\sim Y_e^2$ \citep{yahil:83}. This is caused by the contribution of the degenerate electrons to the pressure before nuclear densities are reached. The leading-order effect of rotation is contained in the ratio $T/|W|$ of the rotational kinetic energy $T$ to the potential binding energy $W$. The linear dependence on $T/|W|$ remains valid for $T/|W| \lesssim 0.09 $. For larger $T/|W|$, the centrifugal support slows the dynamics, leading to weaker dependence on $T/|W|$ of the GW amplitude \citep[e.g.,][]{dimmelmeier:08}. 

Figure \ref{fig:Waveforms} shows the GW strain as a function of time for four select EOSs (upper panel) and for different $Y_e(\rho)$ profiles for the {\tt SFHo} EOS (lower panel) for models with $T/|W| \approx 0.06$. As we can see, different EOSs and $Y_e(\rho)$ profiles produce GWs with amplitudes that differ by $\lesssim 20 \%$ around bounce time. In the post-bounce phase, the differences are more subtle. For a more detailed analysis of the correlation between GW features and the EOS parameters, see \citet{richers:17}.

In the following, we explore if the machine learning model can exploit these differences and classify the EOSs based on the GW signal. We divide our discussion into four parts, in which we separately explore the dependence on the signal range, the number of EOSs in the dataset, the impact of the $Y_e(\rho)$ profiles, and the rotation rate. 

\subsection{Dependence on signal range}
\label{sec:signal_range}

\begin{figure}
    \centering
    \includegraphics[scale=0.5]{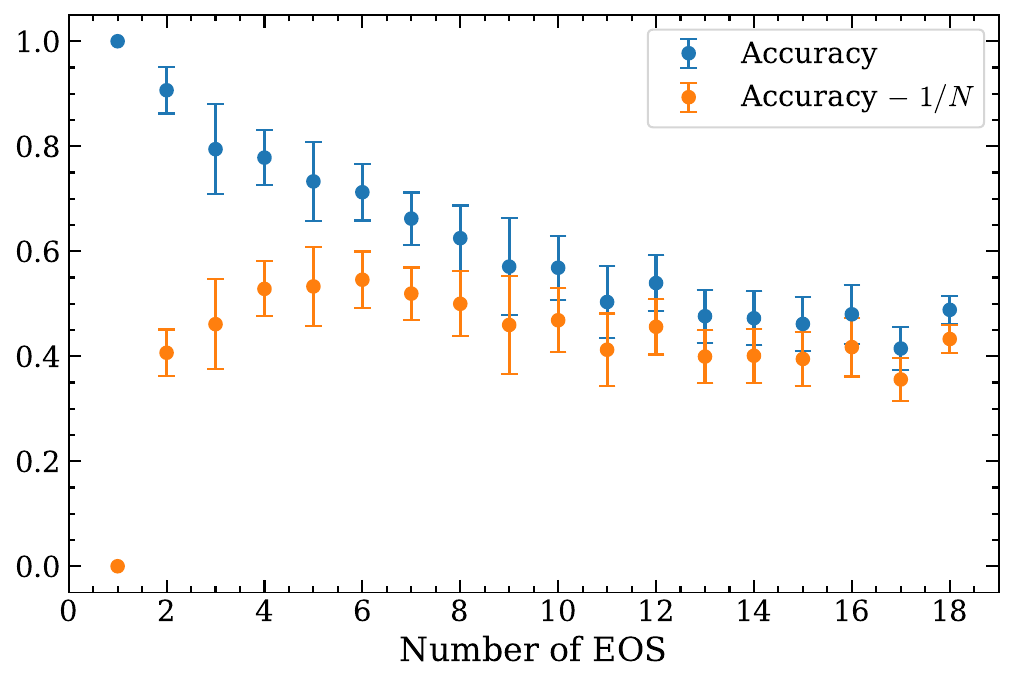}
    \caption{EOS classification accuracy as a function of the number of EOSs in the dataset. The blue points show the mean accuracy. The error bars represent $1 \sigma$ standard deviation. The orange dots are the difference between the CNN classification accuracy and the accuracy of random selection. This quantity measures the advantage that CNN classification offers compared to random selection. 
    } 
    \label{fig:accuracy_check}
\end{figure}

We first perform an analysis of the group 0 dataset that includes GW signals for 18 EOSs. When we perform classification analysis using the GW signal range from -10 ms before bounce to 49 ms after bounce, we obtain an overall accuracy classification accuracy of \(\sim 0.72 \pm 0.07\). This is in agreement with the findings of \citet{Edwards21} on the same dataset. 

However, classification analysis using the GW signal in the range $[-10, 49]$ ms has limitations. First, after $\sim 6$ ms after the bounce, the signal contains contributions of prompt convection \cite[e.g.,][]{dimmelmeier:08}. Since this is a stochastic process, it is hard to capture all possible manifestations of convection with just one simulation per EOS, $Y_e(\rho)$ profile, and progenitor model. Moreover, due to axial symmetry and approximate neutrino treatment used in our simulations, the prompt convection is not modeled accurately. Therefore, the signal after $\sim 6$ ms contains inaccurate features. Second, the GW signal before $-2$ ms has little energy \cite[e.g.,][]{dimmelmeier:08}. Moreover, before $-2$ ms, the inner core density remains below the nuclear density, where various EOSs differ from each other. For these reasons, in the following, we focus on the signal in the $[-2, 6]$ ms range. In this region, the signal is dominated by the core bounce and early ring-down oscillations of the PNS, which is modeled well with the approximations used in our simulations (see Section~\ref{sec:method}). 

For the GW signal in range $[-2, 6]$ ms, we find that the accuracy of classification of 18 EOSs drops to $\sim 0.48 \pm 0.03$, which is significantly lower than that for the $[-10, 49]$ ms range. The reason for this drop is simple: there is less information available in the $[-2, 6]$ ms range compared to that in the $[-10, 49]$ ms range. To support this claim, we calculate the accuracy in the $[6, 49]$ ms range, which we find to be $0.82\pm0.06$. Within the statistical errors, this is comparable to the accuracy in the whole $[-10, 49]$ ms range. This means that there is significant information available in the $[6, 49]$ ms range that the ML model can exploit. As mentioned above, since our model cannot guarantee accuracy in the time frame after $6$ ms, we do not include it in our analysis.

This finding suggests that it is hard to achieve high classification accuracy for a dataset of 18 EOSs based on the bounce GW signal alone. Next, we explore how the classification accuracy depends on the number of EOSs in the dataset. All our results presented hereafter are based on the analysis of the signal in the $ [-2, 6]$ ms range. 

\subsection{Dependence on the number of EOSs}
\label{sec:eos_number}

Figure~\ref{fig:accuracy_check} shows the average classification accuracy as a function of the number of EOSs $N$ ranging from $1$ to $18$. When $N$ is smaller than 18, the results are averaged over 10 random permutations of the 18 EOSs. The blue points show the average accuracy values, while the error bars show the corresponding standard deviation. 

As expected, the accuracy decreases with increasing the number of EOSs. In the region from $N=1$ till $N \sim 11$, the accuracy decreases approximately linearly with $N$, reaching $\sim 0.50 \pm 0.07$ at $N=11$. For larger $N$, the decrease with $N$ is smaller, dropping to $\sim 0.48 \pm 0.03$ for $N = 18$. 

The orange dots in Fig.~\ref{fig:accuracy_check} show the difference between the average classification obtained by the CNN and the accuracy of purely random selection as a function of the number of EOSs. This quantity measures the advantage that CNN classification provides compared to a random selection. As we can see, this quantity reaches a peak value of $\sim 0.55$ at $N=4-7$. At $N \gtrsim 7$, it decreases with $N$, gradually transitioning to its quasi-asymptotic value of $\sim 0.4$. 

This result, in combination with the fact that the classification accuracy decreases with increasing $N$, suggests that at $N=4$ the CNN classification offers the biggest advantage compared to a random selection. For four EOSs, the average CNN accuracy is $0.78 \pm 0.05$. Moreover, $N=4$ is a small enough number of EOSs where we can perform a more in-depth analysis with moderate computational costs, which we do below. At the same time, we emphasize that we are not aware of any other argument in favor using $N=4$ EOSs. We do not claim that all uncertainties in high-density matter properties can be ``grouped" into four distinct EOSs.
 
Hereafter we focus on the classification of four different EOSs. We select {\tt LS220},  {\tt GShenFSU2.1}, {\tt HSDD2}, and {\tt SFHo}. As mentioned above (cf. Section \ref{sec:method}), these four EOSs represent the relatively realistic EOSs that exhibit distinct peak signal frequencies from each other, as can be seen in Fig. 10 of \citet{richers:17}. A similar analysis was performed by \citet{Chao22}. Instead of classifying all 18 EOSs, they grouped the datasets into families of EOSs. In our work, we go beyond \citet{Chao22} by including an analysis of the impact of the uncertainties in the electron fraction and impact of rotation, we as discuss below.

\subsection{Dependence on electron fraction}
\label{sec:ye_dependence}

In this section, we study the performance of the classification algorithm when we add signals that are produced using different electron fraction profiles $Y_e(\rho)$. As mentioned in Section~\ref{sec:method}, we consider three sets of data. Group 1 contains signals generated using the fiducial values of $Y_e(\rho)$, while group 2 contains two extra $Y_e(\rho)$ profiles obtained according to formula (\ref{eq:dye}). Finally, groups 3 and 3b include three more electron fraction profiles (see Section~\ref{sec:method} for details).  

\begin{figure}
    \centering
    \includegraphics[scale=0.42]{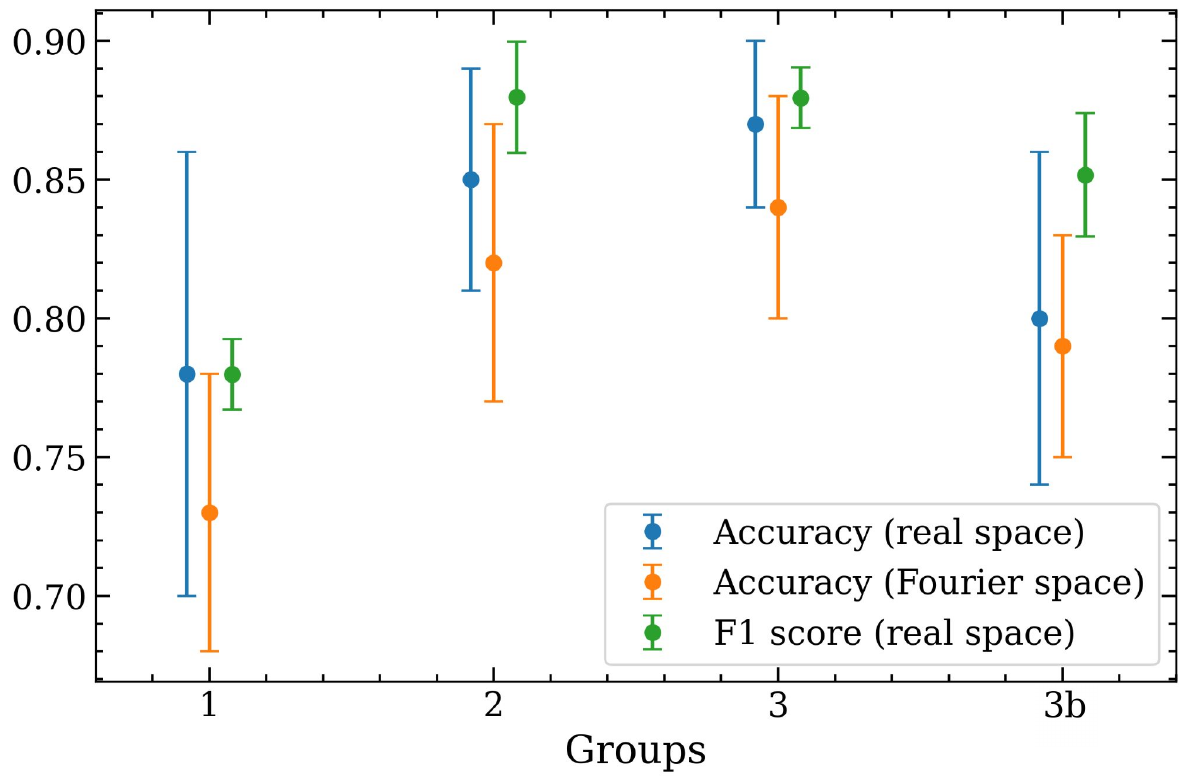}
    \caption{The values of the EOS classification accuracy and F1 score for groups 1, 2, 3, and 3b. The blue dots correspond to the accuracies for the time series data in the Real space, the orange dots represent the accuracies in Fourier space, and the green dots represent F1 scores in Real space.}
    \label{fig:f1}
\end{figure}

Figure~\ref{fig:f1} shows the average classification accuracies for groups 1, 2, 3, and 3b. The accuracies for groups 1, 2, and 3 are \gfirstscore, \gsecscore, and \gthirdscore. These accuracy values can be understood as the result of two opposing factors: the number and complexity of the waveforms contained in each dataset. The former is beneficial to the training of the CNN model, but the latter adversely affects the accuracy. 

The lowest accuracy of \gfirstscore \ exhibited by group 1 is caused by its small sample size of 320, which makes it harder to train the ML model. Group 2 has three times more waveforms, but it also has two more $Y_e(\rho)$ profiles. Nevertheless, group 2 exhibits higher accuracy of \gsecscore. Group 3 has even larger number of 1200 waveforms. The classification accuracy is accordingly the highest. This suggests that the number of waveforms in the dataset is more important to the classification accuracy than the uncertainty in the $Y_e(\rho)$ profiles, at least within the limits considered in this work. 

It is interesting to compare the classification accuracies for groups 3 to 3b. The latter has the same number of $Y_e(\rho)$ profiles, but it contains 240 fewer waveforms. As a result, the group 3b exhibits lower accuracy of \gthirdbscore. This value is lower than the corresponding accuracy for group 2. This is not surprising: group 2 contains the same number of waveforms as group 3b, but it has fewer variations of $Y_e(\rho)$ profiles compared to 3b.

\begin{figure*}
    \centering
    \begin{tabular}{cc}
        \includegraphics[scale=0.27]{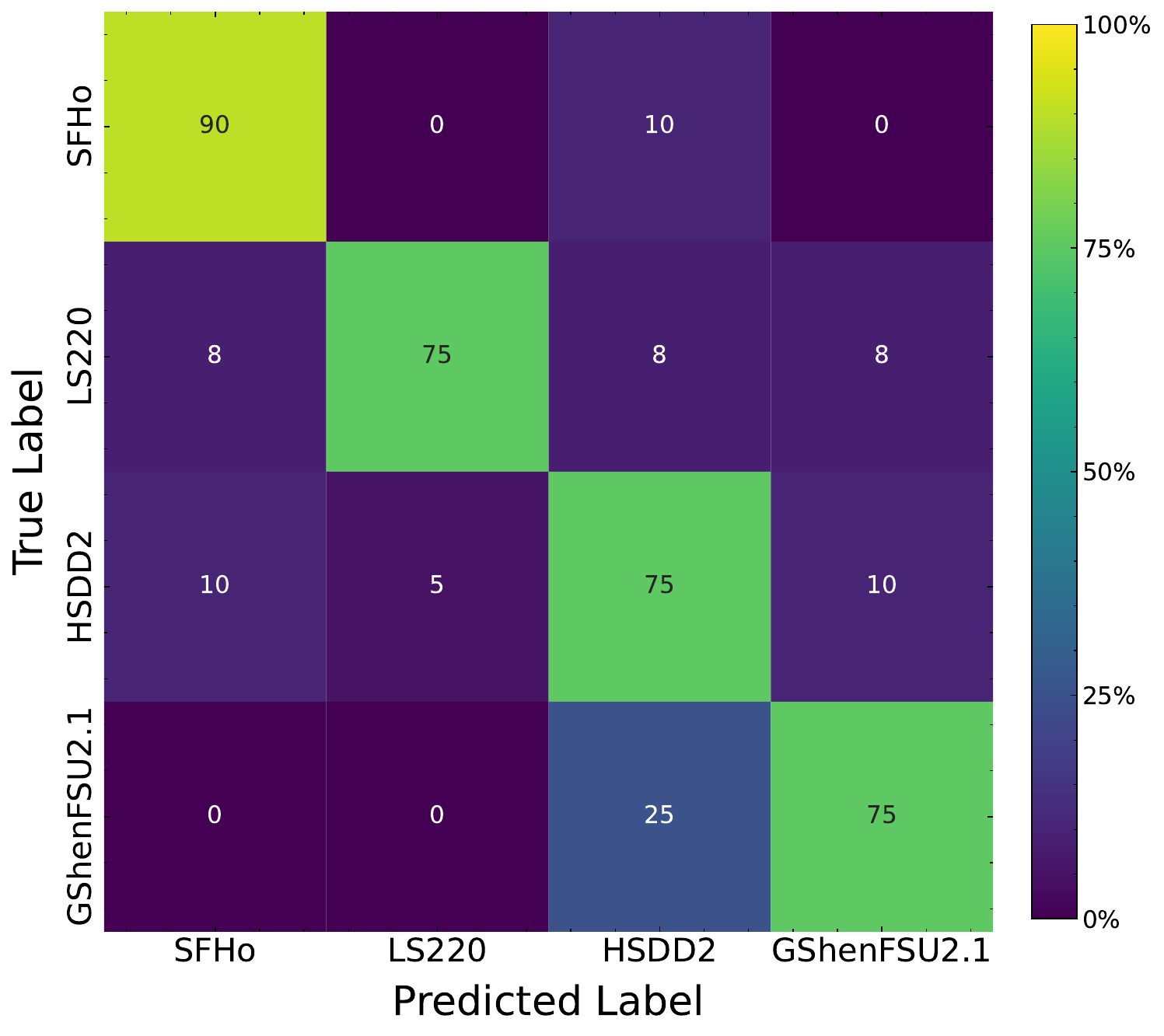} & \includegraphics[scale=0.27]{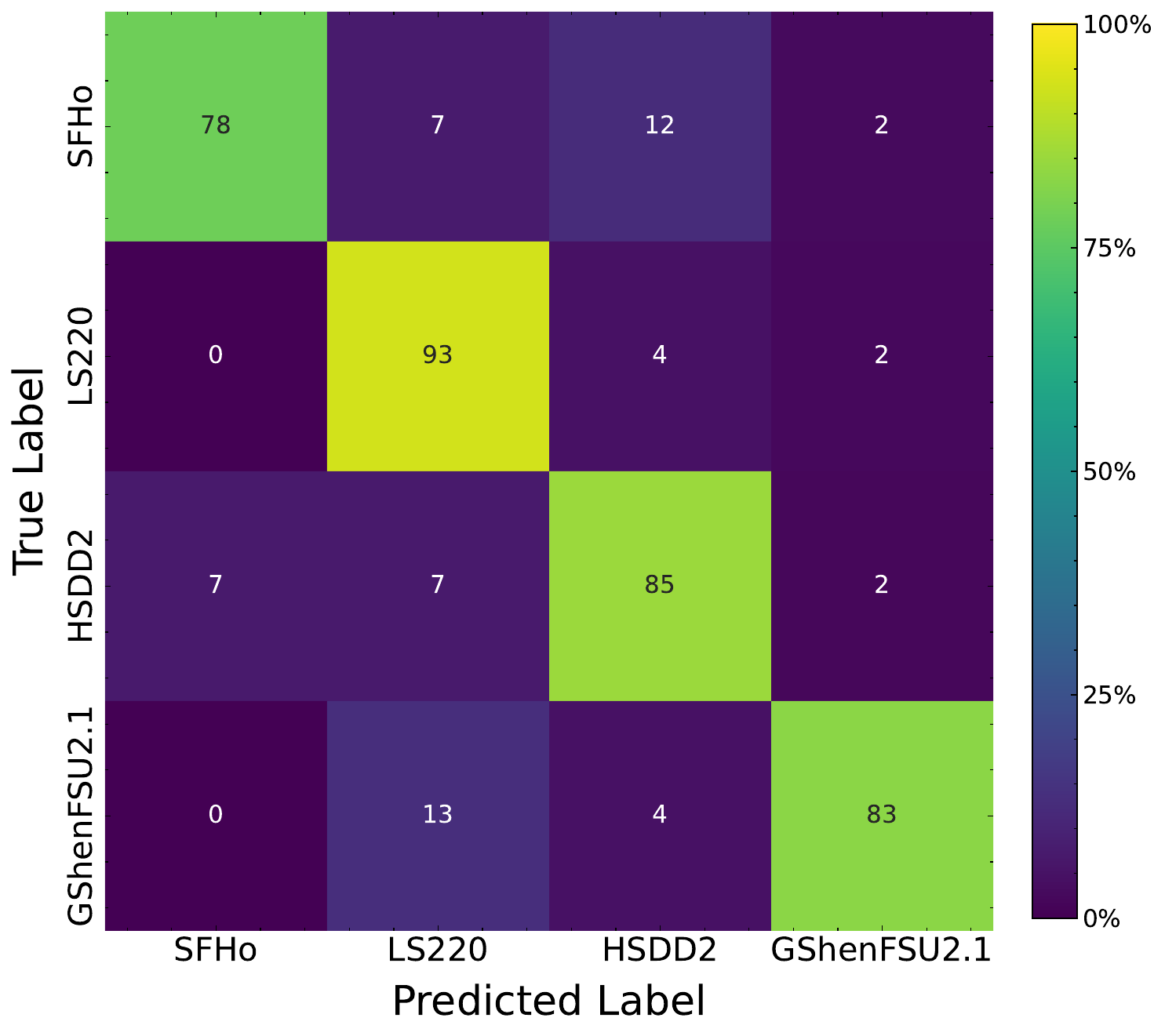} \\
        \includegraphics[scale=0.27]{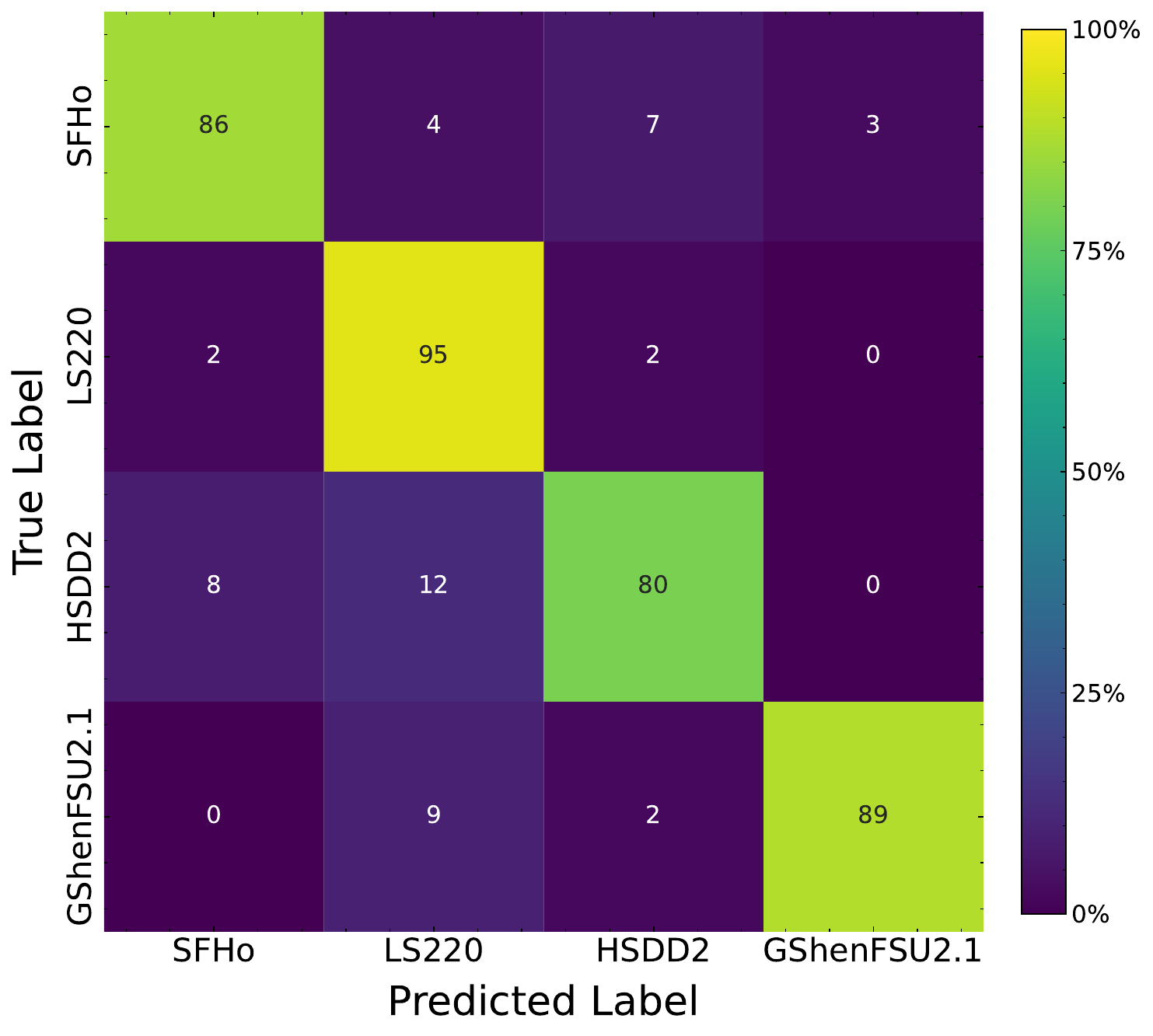} & \includegraphics[scale=0.27]{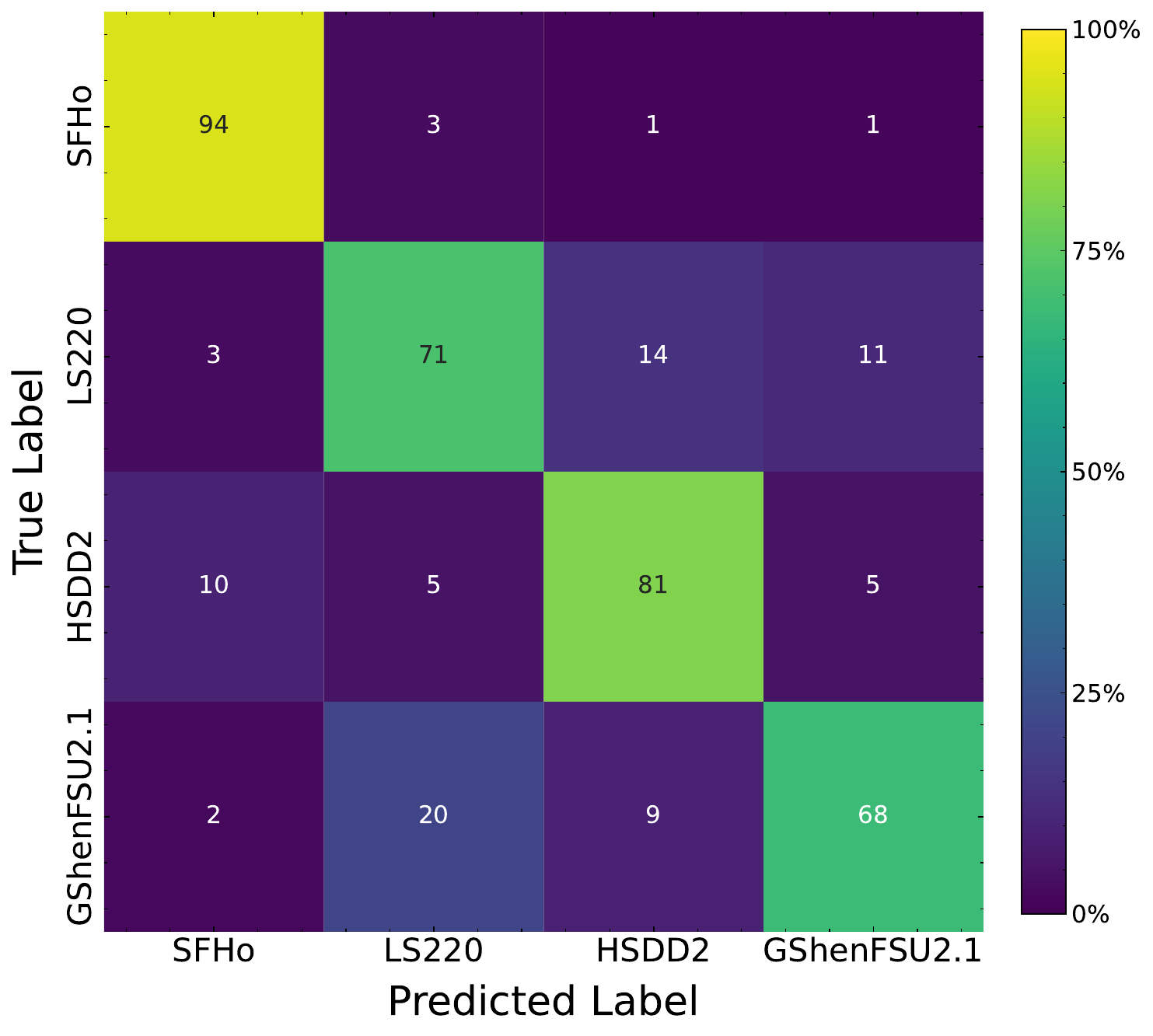} \\
    \end{tabular}
    \caption{Top-left, top-right, bottom-left, and bottom-right panels show the confusion matrices for 4-EOS classification for groups 1, 2, 3, and 3b, respectively. The plots presented here correspond to results from a single run. The respective overall classification are \gfirstscore, \gsecscore, \gthirdscore, and \gthirdbscore.}
    \label{fig:4_eos_CM_group_a}
\end{figure*}

Fig.~\ref{fig:4_eos_CM_group_a} shows confusion matrices from one classification run for groups 1, 2, 3, and 3b. Close inspection do not reveal any trends in terms of classes being misclassified. If we look at the off-diagonal elements, we see that there are no pairs of EOS which are misclassified by the model systematically across all groups. This suggest that the evaluation is not biased systematically. 

For example, in group 1, 25\% of GShenFSU2.1 waveforms were misclassified as HSDD2, while in group 2, only 4\% of GShenFSU2.1 signal were misclassified as HSDD2, and 13\% were misclassified as LS220 data. In both group 1 and 2, 10\% and 12\% of SFHo signal were misclassified with HSDD2, but in groups 3 and 3b significantly fewer misclassifications of LS220 was related to HSDD2. Other EOSs also do not show a misclassification pattern during experiments, which suggests that the misclassification is random.

For additional insight, we look at the F1 score. The F1 score 
can give additional insight over the accuracy metric by providing a measure of the robustness of a model performance, especially for imbalanced datasets. Fig.~\ref{fig:f1} shows the F1 score with the corresponding accuracy score for each group. As we can see, the F1 score shows a similar values to the accuracy score as well as a similar trend across data groups. However, it is important to note that the differences in the F1 scores between groups 2, 3, and 3b are within the standard deviation, so the trends observed in the F1 score (and accuracy) among these groups may not be statistically significant.

To complement the time-series analysis performed so far, we have repeated this analysis in the Fourier space. The corresponding accuracies are shown in Fig.~\ref{fig:f1}. We obtain a similar hierarchy of accuracy values for groups 1, 2, 3, and 3b to that obtained from the time series calculation. However, the accuracies are on average $\sim 3$ percent lower in the Fourier space. A similar drop was observed by \citet{Edwards21}. See Appendix \ref{sec:fourier} for more detailed discussion.

\subsection{Dependence on rotation}
\label{sec:rotation}

\begin{figure}
    \centering
    \includegraphics[scale=0.49]{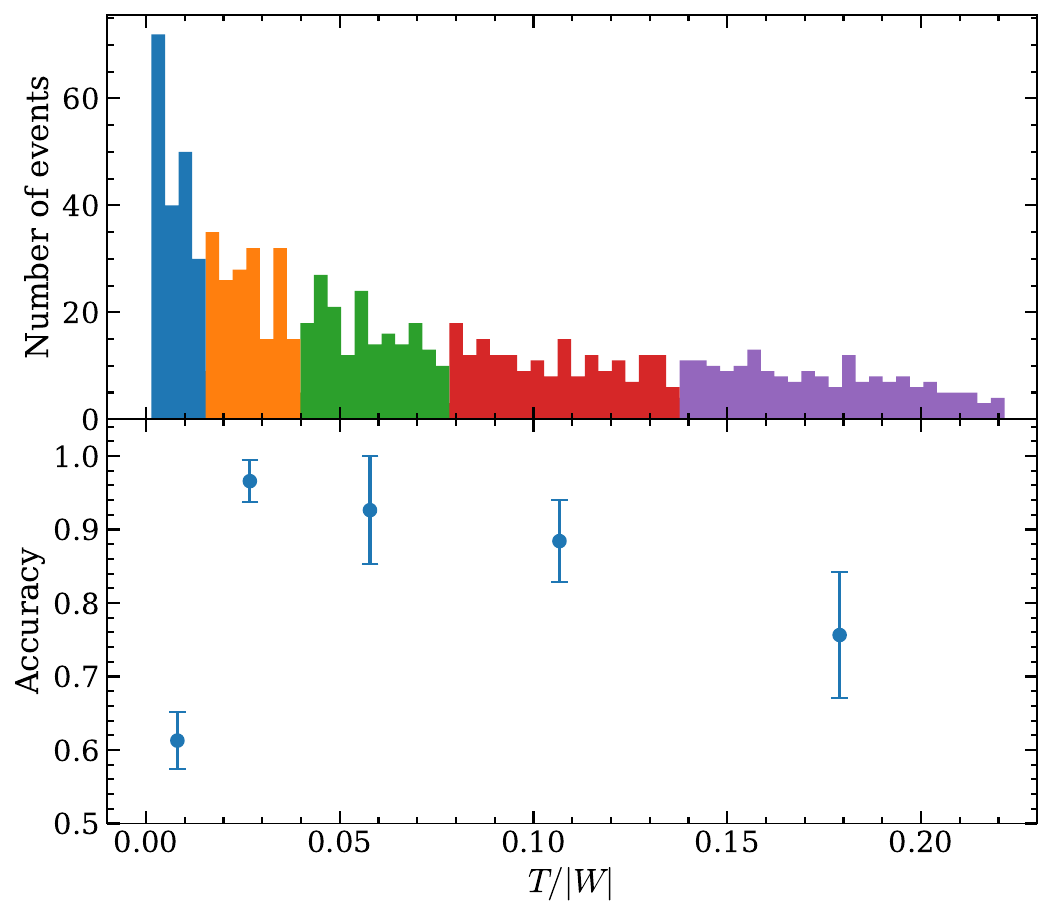}
    \caption{Upper panel: Histogram of $T/|W|$ values and the $5$ bins shown in $5$ different colors. Lower panel: Accuracy of EOS classification in different $T/|W|$ bins. The blue points correspond to the accuracy in the corresponding $T/|W|$ bin. The error bars represent the $1 \sigma$ standard deviation level. The corresponding values are provided in Table \ref{tab:quantiles}. 
    } 
    \label{fig:tw_3b}
\end{figure}

In this section, we study the dependence of the EOS classification accuracy on rotation. We measure rotation in terms of parameter $T/|W|$. We use group 2 dataset, since it has equal number of signals for each EOS and $Y_e$ profile and a bigger size than group 1. 

We group the dataset into five $T/|W|$ bins using quantile cuts, as shown in the upper panel of Fig.~\ref{fig:tw_3b}. For each bin, we compute the classification accuracy. The corresponding accuracies with error bars for each bin are shown in the lower panel of Fig.~\ref{fig:tw_3b}. The accuracy is low for the small and large values of $T/|W|$. For the $ T/|W| < 0.0148$ bin, the average accuracy is $\sim 0.61 \pm 0.04$, while for the $T/|W| > 0.1365$ bin, the average accuracy is $\sim 0.76 \pm 0.09$. This behavior is expected for two reasons. At low $T/|W|$, the signal is weak  (cf. Eq. \ref{eq:hestimate}) and is dominated by numerical noise \cite[e.g.,][]{dimmelmeier:08}. At high $T/|W|$, the centrifugal support prevents PNS from reaching high densities where the differences between EOSs are most pronounced. For example, for the {\tt LS220} EOS, we find that the central density at bounce scales as $\sim 4.4 - 8.5 \times T/|W| \, \mathrm{g/cm^3}$ (cf. Fig. 5 of \cite{abdikamalov:14}). For these reasons, we obtain relatively high classification accuracy for $ 0.0148 < T/|W| < 0.1365$. In this region, rotation is sufficiently strong to induce significant quadrupolar deformation of the inner core. At the same time, centrifugal support is not too strong to prevent PNS from reaching high densities in the center. This is supported by Figs. 6, 8, and 10 of \citet{richers:17}, which reveal that there are significant differences in the GWs corresponding to distinct EOSs even at high $T/|W|$. Table \ref{tab:quantiles} summarizes all the quantitative findings of this section. 

\begin{table}
\centering
\begin{tabular}{c c c c}
\hline
Bin & Bin & Event & Classification \\
No. & Range & Counts & Accuracy  \\
\hline
1 & 0.0014 - 0.0148 & 39 & 0.61 $\pm$ \ 0.04  \\
2 & 0.0148 - 0.0388 & 38 & 0.97 $\pm$ \ 0.03   \\
3 & 0.0388 - 0.0770 & 38 & 0.93 $\pm$\  0.07  \\
4 & 0.0770 - 0.1365 & 38 & 0.88 $\pm$\  0.06  \\
5 & 0.1365 - 0.2216 & 39 & 0.76 $\pm$\  0.09 \\
\hline
\hline
\end{tabular}
\caption{Summary of the EOS classification test in five different bins of the values of rotation parameter $T/|W|$. 
}
\label{tab:quantiles}
\end{table}

%%%%%%%%%%%%%%%%%%%%%%%%%%%%%%%%%%%%%%%%%%%%%%%%%%
\section{Conclusion}
\label{sec:conclusion}

We performed machine learning classification analysis of the nuclear equation of state (EOS) using supernova gravitational wave (GW) signals. We use bounce and early post-bounce GW signal from rotating core-collapse supernovae. We parametrize the uncertainty in the electron capture rates by generating waveforms corresponding to six different electron fraction profiles. The GW signals are obtained from general relativistic hydrodynamics simulations using the {\tt CoCoNuT} code using the $Y_e(\rho)$ deleptonization scheme \citep{liebendoerfer:05b}. For each EOS and $Y_e(\rho)$ model, we perform 98 simulations for different rotational configurations of a $12 M_\odot$ progenitor model. We utilized the 1D CNN model, which was constructed closely following the architecture detailed by \citet{Edwards21} (see Section~\ref{sec:method} for the details of our method). 

We first explore the dependence of the EOS classification accuracy on the GW signal range included in the training and testing of the CNN model. For this, we used the 18 EOS dataset of \citet{richers:17}. The classification accuracy is \(\sim 0.72 \pm 0.07\) for the signal range from $-10$ to $49$ ms, where the origin of the time axis corresponds to the time of bounce. The accuracy decreases gradually with the narrowing of the signal, reaching \(\sim 0.48 \pm 0.03\) for the signal in range from $-2$ to $6$ ms. This range includes only the bounce and early ring-down oscillation signal (see Section \ref{sec:signal_range} for details).

We then study how the accuracy depends on the number of EOSs $N$ included in the data. The accuracy decreases gradually with $N$. However, the difference between the CNN classification accuracy and the accuracy of a random selection exhibits a peak in the region $4 \lesssim N \lesssim 7$. This means that for these values of $N$, the CNN classification offers the greatest advantage compared to random selection (see Section \ref{sec:eos_number} for details). 

Based on this, we then focus on classification analysis of a set of four EOSs: {\tt LS220},  {\tt GShenFSU2.1}, {\tt HSDD2}, and {\tt SFHo}. These EOSs represent relatively realistic EOSs from the set of \citet{richers:17}. At the same time, these EOSs yield relatively distinct peak signal frequencies from each other. This dataset contains 320 waveforms. In this case, the classification accuracy is \gfirstscore (see Section \ref{sec:signal_range} for details). 

Next, we incorporate additional $Y_e(\rho)$ profiles into our dataset. We first add the two different $Y_e(\rho)$ profiles given by Eq.~(\ref{eq:dye}). The dataset size becomes $960$ waveforms. The accuracy increases to \gsecscore. We then add three more $Y_e(\rho)$ profiles and augment the dataset size to 1200 waveforms. In this case, the classification accuracy becomes \gthirdscore. These results suggest that the classification accuracy increases with the dataset size, even if the dataset contains waveforms that are obtained using different $Y_e(\rho)$ profiles (see Section \ref{sec:ye_dependence} for details). 

The classification accuracies weakly depend on rotation. Models with moderately rapid rotation (with $ 0.015 \lesssim T/|W| \lesssim 0.14$) exhibit accuracies $\sim 0.9$. Models with slow ($T/|W| \lesssim 0.015$) and extremely rapid rotation ($T/|W| \gtrsim 0.13 $) have accuracies below $\sim 0.8$. This can be explained by the fact that slow models emit weak GWs, while rapidly rotating models, due to centrifugal support, do not reach high densities, where EOSs differ from each other (see Section \ref{sec:rotation} for details). 

Overall, our work shows the potential of the ML models in inferring the EOS from the GW signal alone, at least for bounce signal in rapidly rotating stars. This is especially true if the selection takes place for a group of four EOSs. However, whether all uncertainties in the parameters of high-density matter can be ``grouped" in family of four EOSs is not clear. For this reason, it is premature to draw definitive conclusions regarding the likelihood of constraining the parameters of nuclear physics using our ML model. Ultimately, instead of classification analysis of EOSs (which is still an insightful exercise), one has to perform a regression analysis of the nuclear parameters from the GW signal. This will be a subject of a future work.   

Our work can be further improved in several other directions. While the assumption of axial symmetry imposed in our simulations is sufficient for the bounce and early post-bounce phase, the subsequent phase requires full 3D modeling \cite[e.g.,][for a recent review]{mueller:20review}. In addition, the simple deleptonization method that we employ cannot capture the complex neutrino process that take place in the post-bounce \cite[e.g.,][]{lentz:12a, kotake2018impact}. Moreover, in our analysis of GWs, we do not include the detector noise that will be present when real detection takes place \cite[e.g.,][]{edwards2017bayesian, bruel2023inference}. Also, there exists room to explore alternative ML algorithms as well as even larger GW datasets. These limitations will be addressed in future works.

%%%%%%%%%%%%%%%%%%%%%%%%%%%%%%%%%%%%%%%%%%%%%%%%%%
\section{Acknowledgements}

We thank Evan O'Connor, Pablo Cerd\'a-Dur\'an, and Jos\'e A.~Font for useful comments. This research is supported by the Science Committee of the Ministry of Science and Higher Education of the Republic of Kazakhstan (Grant No. AP13067834 and AP19677351) and the Nazarbayev University Faculty Development Competitive Research Grant Program (No 11022021FD2912). BS acknowledges funding the Aerospace Committee of the Ministry of Digital Development, Innovations and Aerospace Industry of the Republic of Kazakhstan (Grant No. BR11265408). The authors thank the Institute of Smart Systems and Artificial Intelligence at Nazarbayev University for granting access to the DGX servers.

\section{Data Availability}

The data used in this work is available from authors upon request.

%%%%%%%%%%%%%%%%%%%% REFERENCES %%%%%%%%%%%%%%%%%%

\bibliographystyle{mnras}
\bibliography{references} % if your bibtex file is called example.bib

\begin{thebibliography}{}
\makeatletter
\relax
\def\mn@urlcharsother{\let\do\@makeother \do\$\do\&\do\#\do\^\do\_\do\%\do\~}
\def\mn@doi{\begingroup\mn@urlcharsother \@ifnextchar [ {\mn@doi@} {\mn@doi@[]}}
\def\mn@doi@[#1]#2{\def\@tempa{#1}\ifx\@tempa\@empty \href {http://dx.doi.org/#2} {doi:#2}\else \href {http://dx.doi.org/#2} {#1}\fi \endgroup}
\def\mn@eprint#1#2{\mn@eprint@#1:#2::\@nil}
\def\mn@eprint@arXiv#1{\href {http://arxiv.org/abs/#1} {{\tt arXiv:#1}}}
\def\mn@eprint@dblp#1{\href {http://dblp.uni-trier.de/rec/bibtex/#1.xml} {dblp:#1}}
\def\mn@eprint@#1:#2:#3:#4\@nil{\def\@tempa {#1}\def\@tempb {#2}\def\@tempc {#3}\ifx \@tempc \@empty \let \@tempc \@tempb \let \@tempb \@tempa \fi \ifx \@tempb \@empty \def\@tempb {arXiv}\fi \@ifundefined {mn@eprint@\@tempb}{\@tempb:\@tempc}{\expandafter \expandafter \csname mn@eprint@\@tempb\endcsname \expandafter{\@tempc}}}

\bibitem[\protect\citeauthoryear{Aasi et~al.}{Aasi et~al.}{2015}]{aasi2015advanced}
Aasi J.,  et~al., 2015, Classical and Quantum Gravity, 32, 074001

\bibitem[\protect\citeauthoryear{Abadi et~al.,}{Abadi et~al.}{2016}]{abadi2016tensorflow}
Abadi M.,  et~al., 2016, in 12th USENIX Symposium on Operating Systems Design and Implementation (OSDI 16). pp 265--283

\bibitem[\protect\citeauthoryear{{Abbott} et~al.,}{{Abbott} et~al.}{2016}]{abbot16PRL}
{Abbott} B.~P.,  et~al., 2016, \mn@doi [\prl] {10.1103/PhysRevLett.116.061102}, \href {https://ui.adsabs.harvard.edu/abs/2016PhRvL.116f1102A} {116, 061102}

\bibitem[\protect\citeauthoryear{{Abbott} et~al.,}{{Abbott} et~al.}{2020}]{abbott:20ccsn}
{Abbott} B.~P.,  et~al., 2020, \mn@doi [\prd] {10.1103/PhysRevD.101.084002}, \href {https://ui.adsabs.harvard.edu/abs/2020PhRvD.101h4002A} {101, 084002}

\bibitem[\protect\citeauthoryear{{Abdikamalov}, {Gossan}, {DeMaio}  \& {Ott}}{{Abdikamalov} et~al.}{2014}]{abdikamalov:14}
{Abdikamalov} E.,  {Gossan} S.,  {DeMaio} A.~M.,   {Ott} C.~D.,  2014, \mn@doi [\prd] {10.1103/PhysRevD.90.044001}, \href {https://ui.adsabs.harvard.edu/abs/2014PhRvD..90d4001A} {90, 044001}

\bibitem[\protect\citeauthoryear{Abdikamalov, Pagliaroli  \& Radice}{Abdikamalov et~al.}{2022}]{abdikamalov:22review}
Abdikamalov E.,  Pagliaroli G.,   Radice D.,  2022, in Bambi C.,  Katsanevas S.,   Kokkotas K.~D.,  eds, , Handbook of Gravitational Wave Astronomy.
Springer Singapore, Singapore, p.~21, \mn@doi{10.1007/978-981-15-4702-7_21-1}

\bibitem[\protect\citeauthoryear{Adams, Kochanek, Beacom, Vagins  \& Stanek}{Adams et~al.}{2013}]{adams2013observing}
Adams S.~M.,  Kochanek C.,  Beacom J.~F.,  Vagins M.~R.,   Stanek K.,  2013, The Astrophysical Journal, 778, 164

\bibitem[\protect\citeauthoryear{{Afle} \& {Brown}}{{Afle} \& {Brown}}{2021}]{Afle21}
{Afle} C.,  {Brown} D.~A.,  2021, \mn@doi [\prd] {10.1103/PhysRevD.103.023005}, \href {https://ui.adsabs.harvard.edu/abs/2021PhRvD.103b3005A} {103, 023005}

\bibitem[\protect\citeauthoryear{{Afle}, {Kundu}, {Cammerino}, {Coughlin}, {Brown}, {Vartanyan}  \& {Burrows}}{{Afle} et~al.}{2023}]{Afle23}
{Afle} C.,  {Kundu} S.~K.,  {Cammerino} J.,  {Coughlin} E.~R.,  {Brown} D.~A.,  {Vartanyan} D.,   {Burrows} A.,  2023, \mn@doi [\prd] {10.1103/PhysRevD.107.123005}, \href {https://ui.adsabs.harvard.edu/abs/2023PhRvD.107l3005A} {107, 123005}

\bibitem[\protect\citeauthoryear{Allen, Anderson, Brady, Brown  \& Creighton}{Allen et~al.}{1998}]{allen1998data}
Allen B.,  Anderson W.~G.,  Brady P.~R.,  Brown D.~A.,   Creighton J.~D.,  1998, Physical Review D, 58, 062002

\bibitem[\protect\citeauthoryear{{Andersen}, {Zha}, {da Silva Schneider}, {Betranhandy}, {Couch}  \& {O'Connor}}{{Andersen} et~al.}{2021}]{Andersen21}
{Andersen} O.~E.,  {Zha} S.,  {da Silva Schneider} A.,  {Betranhandy} A.,  {Couch} S.~M.,   {O'Connor} E.~P.,  2021, \mn@doi [\apj] {10.3847/1538-4357/ac294c}, \href {https://ui.adsabs.harvard.edu/abs/2021ApJ...923..201A} {923, 201}

\bibitem[\protect\citeauthoryear{{Andresen}, {M{\"u}ller}, {M{\"u}ller}  \& {Janka}}{{Andresen} et~al.}{2017}]{andresen:17}
{Andresen} H.,  {M{\"u}ller} B.,  {M{\"u}ller} E.,   {Janka} H.~T.,  2017, \mn@doi [\mnras] {10.1093/mnras/stx618}, \href {https://ui.adsabs.harvard.edu/abs/2017MNRAS.468.2032A} {468, 2032}

\bibitem[\protect\citeauthoryear{{Antelis}, {Cavaglia}, {Hansen}, {Morales}, {Moreno}, {Mukherjee}, {Szczepa{\'n}czyk}  \& {Zanolin}}{{Antelis} et~al.}{2022}]{Antelis22}
{Antelis} J.~M.,  {Cavaglia} M.,  {Hansen} T.,  {Morales} M.~D.,  {Moreno} C.,  {Mukherjee} S.,  {Szczepa{\'n}czyk} M.~J.,   {Zanolin} M.,  2022, \mn@doi [\prd] {10.1103/PhysRevD.105.084054}, \href {https://ui.adsabs.harvard.edu/abs/2022PhRvD.105h4054A} {105, 084054}

\bibitem[\protect\citeauthoryear{Baiotti \& Rezzolla}{Baiotti \& Rezzolla}{2017}]{baiotti2017binary}
Baiotti L.,  Rezzolla L.,  2017, Reports on Progress in Physics, 80, 096901

\bibitem[\protect\citeauthoryear{Bauswein et~al.,}{Bauswein et~al.}{2020}]{bauswein2020equation}
Bauswein A.,  et~al., 2020, Physical Review Letters, 125, 141103

\bibitem[\protect\citeauthoryear{Blackman \& Tukey}{Blackman \& Tukey}{1958}]{blackman1958measurement}
Blackman R.,  Tukey J.,  1958, The measurement of power spectra from the point of view of communications engineering.
Dover Publications

\bibitem[\protect\citeauthoryear{{Blondin}, {Mezzacappa}  \& {DeMarino}}{{Blondin} et~al.}{2003}]{blondin:03}
{Blondin} J.~M.,  {Mezzacappa} A.,   {DeMarino} C.,  2003, \mn@doi [\apj] {10.1086/345812}, 584, 971

\bibitem[\protect\citeauthoryear{Bloomfield}{Bloomfield}{2004}]{Bloomfield2004}
Bloomfield P.,  2004, Fourier Analysis of Time Series: An Introduction.
Wiley, New York, NY

\bibitem[\protect\citeauthoryear{Bruel et~al.,}{Bruel et~al.}{2023}]{bruel2023inference}
Bruel T.,  et~al., 2023, Physical Review D, 107, 083029

\bibitem[\protect\citeauthoryear{{Burrows} \& {Vartanyan}}{{Burrows} \& {Vartanyan}}{2021}]{Burrows21review}
{Burrows} A.,  {Vartanyan} D.,  2021, \mn@doi [\nat] {10.1038/s41586-020-03059-w}, \href {https://ui.adsabs.harvard.edu/abs/2021Natur.589...29B} {589, 29}

\bibitem[\protect\citeauthoryear{{Burrows}, {Hayes}  \& {Fryxell}}{{Burrows} et~al.}{1995}]{bhf:95}
{Burrows} A.,  {Hayes} J.,   {Fryxell} B.~A.,  1995, \apj, 450, 830

\bibitem[\protect\citeauthoryear{{Burrows}, {Dessart}, {Livne}, {Ott}  \& {Murphy}}{{Burrows} et~al.}{2007}]{burrows:07b}
{Burrows} A.,  {Dessart} L.,  {Livne} E.,  {Ott} C.~D.,   {Murphy} J.,  2007, \mn@doi [\apj] {10.1086/519161}, \href {http://adsabs.harvard.edu/abs/2007ApJ...664..416B} {664, 416}

\bibitem[\protect\citeauthoryear{Carson, Steiner  \& Yagi}{Carson et~al.}{2019}]{carson2019future}
Carson Z.,  Steiner A.~W.,   Yagi K.,  2019, Physical Review D, 100, 023012

\bibitem[\protect\citeauthoryear{Casallas-Lagos, Antelis, Moreno, Zanolin, Mezzacappa  \& Szczepa{\'n}czyk}{Casallas-Lagos et~al.}{2023}]{casallas2023characterizing}
Casallas-Lagos A.,  Antelis J.~M.,  Moreno C.,  Zanolin M.,  Mezzacappa A.,   Szczepa{\'n}czyk M.~J.,  2023, Physical Review D, 108, 084027

\bibitem[\protect\citeauthoryear{{Chan}, {Heng}  \& {Messenger}}{{Chan} et~al.}{2020}]{Chan20}
{Chan} M.~L.,  {Heng} I.~S.,   {Messenger} C.,  2020, \mn@doi [\prd] {10.1103/PhysRevD.102.043022}, \href {https://ui.adsabs.harvard.edu/abs/2020PhRvD.102d3022C} {102, 043022}

\bibitem[\protect\citeauthoryear{{Chao}, {Su}, {Chen}, {Wang}  \& {Pan}}{{Chao} et~al.}{2022}]{Chao22}
{Chao} Y.-S.,  {Su} C.-Z.,  {Chen} T.-Y.,  {Wang} D.-W.,   {Pan} K.-C.,  2022, \mn@doi [\apj] {10.3847/1538-4357/ac930e}, \href {https://ui.adsabs.harvard.edu/abs/2022ApJ...939...13C} {939, 13}

\bibitem[\protect\citeauthoryear{{Chen}, {Ye}, {Chen}, {Zhao}  \& {Heng}}{{Chen} et~al.}{2020}]{2020arXiv201204193C}
{Chen} P.,  {Ye} J.,  {Chen} G.,  {Zhao} J.,   {Heng} P.-A.,  2020, \mn@doi [arXiv e-prints] {10.48550/arXiv.2012.04193}, \href {https://ui.adsabs.harvard.edu/abs/2020arXiv201204193C} {p. arXiv:2012.04193}

\bibitem[\protect\citeauthoryear{Davide \& Giuseppe}{Davide \& Giuseppe}{2020}]{1361981469291709056}
Davide C.,  Giuseppe J.,  2020, \mn@doi [BMC Genomics] {10.1186/s12864-019-6413-7}, 21, 1

\bibitem[\protect\citeauthoryear{{Deheuvels} et~al.,}{{Deheuvels} et~al.}{2014}]{deheuvels:14}
{Deheuvels} S.,  et~al., 2014, \mn@doi [\aap] {10.1051/0004-6361/201322779}, \href {https://ui.adsabs.harvard.edu/abs/2014A&A...564A..27D} {564, A27}

\bibitem[\protect\citeauthoryear{{Dimmelmeier}, {Novak}, {Font}, {Ib{\'a}{\~n}ez}  \& {M{\"u}ller}}{{Dimmelmeier} et~al.}{2005}]{dimmelmeier:05MdM}
{Dimmelmeier} H.,  {Novak} J.,  {Font} J.~A.,  {Ib{\'a}{\~n}ez} J.~M.,   {M{\"u}ller} E.,  2005, \mn@doi [Phys.\ Rev.\ D] {10.1103/PhysRevD.71.064023}, \href {http://adsabs.harvard.edu/cgi-bin/nph-bib_query?bibcode=2005PhRvD..71f4023D&db_key=AST} {71, 064023}

\bibitem[\protect\citeauthoryear{{Dimmelmeier}, {Ott}, {Marek}  \& {Janka}}{{Dimmelmeier} et~al.}{2008}]{dimmelmeier:08}
{Dimmelmeier} H.,  {Ott} C.~D.,  {Marek} A.,   {Janka} H.~T.,  2008, \mn@doi [\prd] {10.1103/PhysRevD.78.064056}, \href {https://ui.adsabs.harvard.edu/abs/2008PhRvD..78f4056D} {78, 064056}

\bibitem[\protect\citeauthoryear{Edwards}{Edwards}{2017}]{edwards2017bayesian}
Edwards M.,  2017, PhD thesis, ResearchSpace@ Auckland

\bibitem[\protect\citeauthoryear{{Edwards}}{{Edwards}}{2021}]{Edwards21}
{Edwards} M.~C.,  2021, \mn@doi [\prd] {10.1103/PhysRevD.103.024025}, \href {https://ui.adsabs.harvard.edu/abs/2021PhRvD.103b4025E} {103, 024025}

\bibitem[\protect\citeauthoryear{{Edwards}, {Meyer}  \& {Christensen}}{{Edwards} et~al.}{2014}]{Edwards14}
{Edwards} M.~C.,  {Meyer} R.,   {Christensen} N.,  2014, \mn@doi [Inverse Problems] {10.1088/0266-5611/30/11/114008}, \href {https://ui.adsabs.harvard.edu/abs/2014InvPr..30k4008E} {30, 114008}

\bibitem[\protect\citeauthoryear{{Endeve}, {Cardall}, {Budiardja}, {Beck}, {Bejnood}, {Toedte}, {Mezzacappa}  \& {Blondin}}{{Endeve} et~al.}{2012}]{endeve:12}
{Endeve} E.,  {Cardall} C.~Y.,  {Budiardja} R.~D.,  {Beck} S.~W.,  {Bejnood} A.,  {Toedte} R.~J.,  {Mezzacappa} A.,   {Blondin} J.~M.,  2012, \mn@doi [\apj] {10.1088/0004-637X/751/1/26}, \href {http://adsabs.harvard.edu/abs/2012ApJ...751...26E} {751, 26}

\bibitem[\protect\citeauthoryear{{Engels}, {Frey}  \& {Ott}}{{Engels} et~al.}{2014}]{Engels14}
{Engels} W.~J.,  {Frey} R.,   {Ott} C.~D.,  2014, \mn@doi [\prd] {10.1103/PhysRevD.90.124026}, \href {https://ui.adsabs.harvard.edu/abs/2014PhRvD..90l4026E} {90, 124026}

\bibitem[\protect\citeauthoryear{{Ertl}, {Janka}, {Woosley}, {Sukhbold}  \& {Ugliano}}{{Ertl} et~al.}{2016}]{ertl:16}
{Ertl} T.,  {Janka} H.-T.,  {Woosley} S.~E.,  {Sukhbold} T.,   {Ugliano} M.,  2016, \mn@doi [\apj] {10.3847/0004-637X/818/2/124}, \href {http://adsabs.harvard.edu/abs/2016ApJ...818..124E} {818, 124}

\bibitem[\protect\citeauthoryear{Fawaz, Forestier, Weber, Idoumghar  \& Muller}{Fawaz et~al.}{2019}]{fawaz2019deep}
Fawaz H.~I.,  Forestier G.,  Weber J.,  Idoumghar L.,   Muller P.-A.,  2019, Data Mining and Knowledge Discovery, 33, 917

\bibitem[\protect\citeauthoryear{{Fern{\'a}ndez}}{{Fern{\'a}ndez}}{2015}]{fernandez:15a}
{Fern{\'a}ndez} R.,  2015, \mn@doi [\mnras] {10.1093/mnras/stv1463}, \href {http://adsabs.harvard.edu/abs/2015MNRAS.452.2071F} {452, 2071}

\bibitem[\protect\citeauthoryear{{Fern{\'a}ndez} \& {Thompson}}{{Fern{\'a}ndez} \& {Thompson}}{2009}]{fernandez:09a}
{Fern{\'a}ndez} R.,  {Thompson} C.,  2009, \apj, \href {http://adsabs.harvard.edu/abs/2009ApJ...697.1827F} {697, 1827}

\bibitem[\protect\citeauthoryear{{Foglizzo}, {Scheck}  \& {Janka}}{{Foglizzo} et~al.}{2006}]{foglizzo:06}
{Foglizzo} T.,  {Scheck} L.,   {Janka} H.-T.,  2006, \mn@doi [\apj] {10.1086/508443}, 652, 1436

\bibitem[\protect\citeauthoryear{{Fryer} \& {Heger}}{{Fryer} \& {Heger}}{2005}]{fryer:05}
{Fryer} C.~L.,  {Heger} A.,  2005, \mn@doi [\apj] {10.1086/428379}, \href {https://ui.adsabs.harvard.edu/abs/2005ApJ...623..302F} {623, 302}

\bibitem[\protect\citeauthoryear{{Fuller}, {Klion}, {Abdikamalov}  \& {Ott}}{{Fuller} et~al.}{2015}]{fuller:15}
{Fuller} J.,  {Klion} H.,  {Abdikamalov} E.,   {Ott} C.~D.,  2015, \mn@doi [\mnras] {10.1093/mnras/stv698}, \href {https://ui.adsabs.harvard.edu/abs/2015MNRAS.450..414F} {450, 414}

\bibitem[\protect\citeauthoryear{George \& Huerta}{George \& Huerta}{2018}]{george2018deep}
George D.,  Huerta E.~A.,  2018, Physical Review D, 97, 044039

\bibitem[\protect\citeauthoryear{Goodfellow, Bengio  \& Courville}{Goodfellow et~al.}{2016}]{goodfellow2016deep}
Goodfellow I.,  Bengio Y.,   Courville A.,  2016, Deep learning.
MIT press

\bibitem[\protect\citeauthoryear{{Gossan}, {Sutton}, {Stuver}, {Zanolin}, {Gill}  \& {Ott}}{{Gossan} et~al.}{2016}]{gossan:16}
{Gossan} S.~E.,  {Sutton} P.,  {Stuver} A.,  {Zanolin} M.,  {Gill} K.,   {Ott} C.~D.,  2016, \mn@doi [\prd] {10.1103/PhysRevD.93.042002}, \href {https://ui.adsabs.harvard.edu/abs/2016PhRvD..93d2002G} {93, 042002}

\bibitem[\protect\citeauthoryear{Hanin}{Hanin}{2019}]{hanin2019universal}
Hanin B.,  2019, Mathematics, 7, 992

\bibitem[\protect\citeauthoryear{{Hayama}, {Kuroda}, {Nakamura}  \& {Yamada}}{{Hayama} et~al.}{2016}]{hayama16}
{Hayama} K.,  {Kuroda} T.,  {Nakamura} K.,   {Yamada} S.,  2016, \mn@doi [\prl] {10.1103/PhysRevLett.116.151102}, \href {https://ui.adsabs.harvard.edu/abs/2016PhRvL.116o1102H} {116, 151102}

\bibitem[\protect\citeauthoryear{He, Zhang, Ren  \& Sun}{He et~al.}{2016}]{he2016deep}
He K.,  Zhang X.,  Ren S.,   Sun J.,  2016, in Proceedings of the IEEE conference on computer vision and pattern recognition. pp 770--778

\bibitem[\protect\citeauthoryear{{Heger}, {Woosley}  \& {Spruit}}{{Heger} et~al.}{2005}]{heger:05}
{Heger} A.,  {Woosley} S.~E.,   {Spruit} H.~C.,  2005, \apj, 626, 350

\bibitem[\protect\citeauthoryear{{Hempel} \& {Schaffner-Bielich}}{{Hempel} \& {Schaffner-Bielich}}{2010}]{hempel:10}
{Hempel} M.,  {Schaffner-Bielich} J.,  2010, \mn@doi [\nphysa] {10.1016/j.nuclphysa.2010.02.010}, \href {http://adsabs.harvard.edu/abs/2010NuPhA.837..210H} {837, 210}

\bibitem[\protect\citeauthoryear{{Hempel}, {Fischer}, {Schaffner-Bielich}  \& {Liebend{\"o}rfer}}{{Hempel} et~al.}{2012}]{hempel:12}
{Hempel} M.,  {Fischer} T.,  {Schaffner-Bielich} J.,   {Liebend{\"o}rfer} M.,  2012, \mn@doi [\apj] {10.1088/0004-637X/748/1/70}, \href {http://adsabs.harvard.edu/abs/2012ApJ...748...70H} {748, 70}

\bibitem[\protect\citeauthoryear{{Herant}, {Benz}, {Hix}, {Fryer}  \& {Colgate}}{{Herant} et~al.}{1994}]{herant:94}
{Herant} M.,  {Benz} W.,  {Hix} W.~R.,  {Fryer} C.~L.,   {Colgate} S.~A.,  1994, \apj, 435, 339

\bibitem[\protect\citeauthoryear{{Hix}, {Messer}, {Mezzacappa}, {Liebend{\"o}rfer}, {Sampaio}, {Langanke}, {Dean}  \& {Mart{\'\i}nez-Pinedo}}{{Hix} et~al.}{2003}]{Hix:03}
{Hix} W.~R.,  {Messer} O.~E.,  {Mezzacappa} A.,  {Liebend{\"o}rfer} M.,  {Sampaio} J.,  {Langanke} K.,  {Dean} D.~J.,   {Mart{\'\i}nez-Pinedo} G.,  2003, \mn@doi [\prl] {10.1103/PhysRevLett.91.201102}, \href {https://ui.adsabs.harvard.edu/abs/2003PhRvL..91t1102H} {91, 201102}

\bibitem[\protect\citeauthoryear{Huang, Liu, Van Der~Maaten  \& Weinberger}{Huang et~al.}{2017}]{huang2017densely}
Huang G.,  Liu Z.,  Van Der~Maaten L.,   Weinberger K.~Q.,  2017, in Proceedings of the IEEE conference on computer vision and pattern recognition. pp 4700--4708

\bibitem[\protect\citeauthoryear{Iacovelli, Mancarella, Mondal, Puecher, Dietrich, Gulminelli, Maggiore  \& Oertel}{Iacovelli et~al.}{2023}]{iacovelli2023nuclear}
Iacovelli F.,  Mancarella M.,  Mondal C.,  Puecher A.,  Dietrich T.,  Gulminelli F.,  Maggiore M.,   Oertel M.,  2023, arXiv preprint arXiv:2308.12378

\bibitem[\protect\citeauthoryear{{Iwakami}, {Kotake}, {Ohnishi}, {Yamada}  \& {Sawada}}{{Iwakami} et~al.}{2009}]{iwakami:09}
{Iwakami} W.,  {Kotake} K.,  {Ohnishi} N.,  {Yamada} S.,   {Sawada} K.,  2009, \apj, \href {http://adsabs.harvard.edu/abs/2009ApJ...700..232I} {700, 232}

\bibitem[\protect\citeauthoryear{{Janka}}{{Janka}}{2001}]{janka:01}
{Janka} H.-T.,  2001, \mn@doi [\aap] {10.1051/0004-6361:20010012}, \href {http://adsabs.harvard.edu/abs/2001A%26A...368..527J} {368, 527}

\bibitem[\protect\citeauthoryear{{Janka} \& {M\"uller}}{{Janka} \& {M\"uller}}{1996}]{janka:96}
{Janka} H.-T.,  {M\"uller} E.,  1996, \aap, 306, 167

\bibitem[\protect\citeauthoryear{{Janka}, {Melson}  \& {Summa}}{{Janka} et~al.}{2016}]{janka:16a}
{Janka} H.-T.,  {Melson} T.,   {Summa} A.,  2016, \mn@doi [Ann. Rev. Nuc. Part. Sc.] {10.1146/annurev-nucl-102115-044747}, \href {http://adsabs.harvard.edu/abs/2016ARNPS..66..341J} {66, 341}

\bibitem[\protect\citeauthoryear{Kingma \& Ba}{Kingma \& Ba}{2014}]{kingma2014adam}
Kingma D.~P.,  Ba J.,  2014, arXiv preprint arXiv:1412.6980

\bibitem[\protect\citeauthoryear{Kotake, Takiwaki, Fischer, Nakamura  \& Mart{\'\i}nez-Pinedo}{Kotake et~al.}{2018}]{kotake2018impact}
Kotake K.,  Takiwaki T.,  Fischer T.,  Nakamura K.,   Mart{\'\i}nez-Pinedo G.,  2018, The Astrophysical Journal, 853, 170

\bibitem[\protect\citeauthoryear{Krizhevsky, Sutskever  \& Hinton}{Krizhevsky et~al.}{2012}]{krizhevsky2012imagenet}
Krizhevsky A.,  Sutskever I.,   Hinton G.~E.,  2012, in Advances in neural information processing systems. pp 1097--1105

\bibitem[\protect\citeauthoryear{{Kuroda}, {Kotake}, {Hayama}  \& {Takiwaki}}{{Kuroda} et~al.}{2017}]{kuroda:17}
{Kuroda} T.,  {Kotake} K.,  {Hayama} K.,   {Takiwaki} T.,  2017, \mn@doi [\apj] {10.3847/1538-4357/aa988d}, \href {http://adsabs.harvard.edu/abs/2017ApJ...851...62K} {851, 62}

\bibitem[\protect\citeauthoryear{{Kuroda}, {Arcones}, {Takiwaki}  \& {Kotake}}{{Kuroda} et~al.}{2020}]{kuroda:20}
{Kuroda} T.,  {Arcones} A.,  {Takiwaki} T.,   {Kotake} K.,  2020, \mn@doi [\apj] {10.3847/1538-4357/ab9308}, \href {https://ui.adsabs.harvard.edu/abs/2020ApJ...896..102K} {896, 102}

\bibitem[\protect\citeauthoryear{Langanke, Mart{\'\i}nez-Pinedo  \& Zegers}{Langanke et~al.}{2021}]{langanke2021electron}
Langanke K.,  Mart{\'\i}nez-Pinedo G.,   Zegers R.,  2021, Reports on Progress in Physics, 84, 066301

\bibitem[\protect\citeauthoryear{Lattimer}{Lattimer}{2023}]{lattimer2023constraints}
Lattimer J.~M.,  2023, in Journal of Physics: Conference Series. p. 012009

\bibitem[\protect\citeauthoryear{{Lattimer} \& {Swesty}}{{Lattimer} \& {Swesty}}{1991}]{lseos:91}
{Lattimer} J.~M.,  {Swesty} F.~D.,  1991, \mn@doi [\nphysa] {10.1016/0375-9474(91)90452-C}, \href {http://adsabs.harvard.edu/abs/1991NuPhA.535..331L} {535, 331}

\bibitem[\protect\citeauthoryear{LeCun, Boser, Denker, Henderson, Howard, Hubbard  \& Jackel}{LeCun et~al.}{1989}]{lecun1989backpropagation}
LeCun Y.,  Boser B.,  Denker J.~S.,  Henderson D.,  Howard R.~E.,  Hubbard W.,   Jackel L.~D.,  1989, Neural computation, 1, 541

\bibitem[\protect\citeauthoryear{LeCun, Bottou, Bengio  \& Haffner}{LeCun et~al.}{1998}]{lecun1998gradient}
LeCun Y.,  Bottou L.,  Bengio Y.,   Haffner P.,  1998, Proceedings of the IEEE, 86, 2278

\bibitem[\protect\citeauthoryear{LeCun, Bengio  \& Hinton}{LeCun et~al.}{2015}]{lecun2015deep}
LeCun Y.,  Bengio Y.,   Hinton G.,  2015, Nature, 521, 436

\bibitem[\protect\citeauthoryear{{Lentz}, {Mezzacappa}, {Messer}, {Liebend{\"o}rfer}, {Hix}  \& {Bruenn}}{{Lentz} et~al.}{2012}]{lentz:12a}
{Lentz} E.~J.,  {Mezzacappa} A.,  {Messer} O.~E.~B.,  {Liebend{\"o}rfer} M.,  {Hix} W.~R.,   {Bruenn} S.~W.,  2012, \mn@doi [\apj] {10.1088/0004-637X/747/1/73}, \href {http://adsabs.harvard.edu/abs/2012ApJ...747...73L} {747, 73}

\bibitem[\protect\citeauthoryear{{Liebend{\"o}rfer}}{{Liebend{\"o}rfer}}{2005}]{liebendoerfer:05b}
{Liebend{\"o}rfer} M.,  2005, \apj, 633, 1042

\bibitem[\protect\citeauthoryear{{Logue}, {Ott}, {Heng}, {Kalmus}  \& {Scargill}}{{Logue} et~al.}{2012}]{Logue12}
{Logue} J.,  {Ott} C.~D.,  {Heng} I.~S.,  {Kalmus} P.,   {Scargill} J.~H.~C.,  2012, \mn@doi [\prd] {10.1103/PhysRevD.86.044023}, \href {https://ui.adsabs.harvard.edu/abs/2012PhRvD..86d4023L} {86, 044023}

\bibitem[\protect\citeauthoryear{{L{\'o}pez}, {Di Palma}, {Drago}, {Cerd{\'a}-Dur{\'a}n}  \& {Ricci}}{{L{\'o}pez} et~al.}{2021}]{Lopez21}
{L{\'o}pez} M.,  {Di Palma} I.,  {Drago} M.,  {Cerd{\'a}-Dur{\'a}n} P.,   {Ricci} F.,  2021, \mn@doi [\prd] {10.1103/PhysRevD.103.063011}, \href {https://ui.adsabs.harvard.edu/abs/2021PhRvD.103f3011L} {103, 063011}

\bibitem[\protect\citeauthoryear{{Mezzacappa} et~al.,}{{Mezzacappa} et~al.}{2023}]{mezzacappa23b}
{Mezzacappa} A.,  et~al., 2023, \mn@doi [\prd] {10.1103/PhysRevD.107.043008}, \href {https://ui.adsabs.harvard.edu/abs/2023PhRvD.107d3008M} {107, 043008}

\bibitem[\protect\citeauthoryear{{Mitra}, {Shukirgaliyev}, {Abylkairov}  \& {Abdikamalov}}{{Mitra} et~al.}{2023}]{mitra23}
{Mitra} A.,  {Shukirgaliyev} B.,  {Abylkairov} Y.~S.,   {Abdikamalov} E.,  2023, \mn@doi [\mnras] {10.1093/mnras/stad169}, \href {https://ui.adsabs.harvard.edu/abs/2023MNRAS.520.2473M} {520, 2473}

\bibitem[\protect\citeauthoryear{{Mori}, {Suwa}  \& {Takiwaki}}{{Mori} et~al.}{2023}]{Mori23}
{Mori} M.,  {Suwa} Y.,   {Takiwaki} T.,  2023, \mn@doi [\prd] {10.1103/PhysRevD.107.083015}, \href {https://ui.adsabs.harvard.edu/abs/2023PhRvD.107h3015M} {107, 083015}

\bibitem[\protect\citeauthoryear{{Morozova}, {Radice}, {Burrows}  \& {Vartanyan}}{{Morozova} et~al.}{2018}]{morozova:18}
{Morozova} V.,  {Radice} D.,  {Burrows} A.,   {Vartanyan} D.,  2018, \mn@doi [\apj] {10.3847/1538-4357/aac5f1}, \href {https://ui.adsabs.harvard.edu/abs/2018ApJ...861...10M} {861, 10}

\bibitem[\protect\citeauthoryear{{Mosser} et~al.,}{{Mosser} et~al.}{2012}]{mosser:12}
{Mosser} B.,  et~al., 2012, \mn@doi [\aap] {10.1051/0004-6361/201220106}, \href {https://ui.adsabs.harvard.edu/abs/2012A&A...548A..10M} {548, A10}

\bibitem[\protect\citeauthoryear{{M{\"o}sta} et~al.,}{{M{\"o}sta} et~al.}{2014}]{moesta:14b}
{M{\"o}sta} P.,  et~al., 2014, \mn@doi [\apjl] {10.1088/2041-8205/785/2/L29}, \href {http://adsabs.harvard.edu/abs/2014ApJ...785L..29M} {785, L29}

\bibitem[\protect\citeauthoryear{{Mueller} \& {Janka}}{{Mueller} \& {Janka}}{1997}]{mueller:97}
{Mueller} E.,  {Janka} H.~T.,  1997, \aap, \href {https://ui.adsabs.harvard.edu/abs/1997A&A...317..140M} {317, 140}

\bibitem[\protect\citeauthoryear{M\"uller}{M\"uller}{2009}]{mueller:09phd}
M\"uller B.,  2009, PhD thesis, {T}echnische {U}niversit\"at {M}\"unchen, {M}\"unchen, {G}ermany

\bibitem[\protect\citeauthoryear{{M{\"u}ller}}{{M{\"u}ller}}{2020}]{mueller:20review}
{M{\"u}ller} B.,  2020, \mn@doi [Living Reviews in Computational Astrophysics] {10.1007/s41115-020-0008-5}, \href {https://ui.adsabs.harvard.edu/abs/2020LRCA....6....3M} {6, 3}

\bibitem[\protect\citeauthoryear{{M{\"u}ller} \& {Janka}}{{M{\"u}ller} \& {Janka}}{2015}]{mueller:15}
{M{\"u}ller} B.,  {Janka} H.-T.,  2015, \mn@doi [\mnras] {10.1093/mnras/stv101}, \href {http://adsabs.harvard.edu/abs/2015MNRAS.448.2141M} {448, 2141}

\bibitem[\protect\citeauthoryear{{M{\"u}ller} \& {Varma}}{{M{\"u}ller} \& {Varma}}{2020}]{mueller:20b}
{M{\"u}ller} B.,  {Varma} V.,  2020, \mn@doi [\mnras] {10.1093/mnrasl/slaa137}, \href {https://ui.adsabs.harvard.edu/abs/2020MNRAS.498L.109M} {498, L109}

\bibitem[\protect\citeauthoryear{{M{\"u}ller}, {Janka}  \& {Heger}}{{M{\"u}ller} et~al.}{2012}]{mueller:12b}
{M{\"u}ller} B.,  {Janka} H.-T.,   {Heger} A.,  2012, \mn@doi [\apj] {10.1088/0004-637X/761/1/72}, \href {http://adsabs.harvard.edu/abs/2012ApJ...761...72M} {761, 72}

\bibitem[\protect\citeauthoryear{{M{\"u}ller}, {Janka}  \& {Marek}}{{M{\"u}ller} et~al.}{2013}]{mueller:13}
{M{\"u}ller} B.,  {Janka} H.-T.,   {Marek} A.,  2013, \mn@doi [\apj] {10.1088/0004-637X/766/1/43}, \href {https://ui.adsabs.harvard.edu/abs/2013ApJ...766...43M} {766, 43}

\bibitem[\protect\citeauthoryear{{Murphy}, {Dolence}  \& {Burrows}}{{Murphy} et~al.}{2013}]{murphy:13}
{Murphy} J.~W.,  {Dolence} J.~C.,   {Burrows} A.,  2013, \mn@doi [\apj] {10.1088/0004-637X/771/1/52}, \href {http://adsabs.harvard.edu/abs/2013ApJ...771...52M} {771, 52}

\bibitem[\protect\citeauthoryear{Nakamura, Horiuchi, Tanaka, Hayama, Takiwaki  \& Kotake}{Nakamura et~al.}{2016}]{Nakamura:2016kkl}
Nakamura K.,  Horiuchi S.,  Tanaka M.,  Hayama K.,  Takiwaki T.,   Kotake K.,  2016, \mn@doi [\mnras] {10.1093/mnras/stw1453}, 461, 3296

\bibitem[\protect\citeauthoryear{{O'Connor}}{{O'Connor}}{2015}]{oconnor:15a}
{O'Connor} E.,  2015, \mn@doi [\apjs] {10.1088/0067-0049/219/2/24}, \href {http://adsabs.harvard.edu/abs/2015ApJS..219...24O} {219, 24}

\bibitem[\protect\citeauthoryear{{Obergaulinger} \& {Aloy}}{{Obergaulinger} \& {Aloy}}{2020}]{obergaulinger:20}
{Obergaulinger} M.,  {Aloy} M.~{\'A}.,  2020, \mn@doi [\mnras] {10.1093/mnras/staa096}, \href {https://ui.adsabs.harvard.edu/abs/2020MNRAS.492.4613O} {492, 4613}

\bibitem[\protect\citeauthoryear{{Obergaulinger}, {Aloy}  \& {M{\"u}ller}}{{Obergaulinger} et~al.}{2006}]{obergaulinger:06}
{Obergaulinger} M.,  {Aloy} M.~A.,   {M{\"u}ller} E.,  2006, \mn@doi [\aap] {10.1051/0004-6361:20054306}, \href {https://ui.adsabs.harvard.edu/abs/2006A&A...450.1107O} {450, 1107}

\bibitem[\protect\citeauthoryear{{Ott}, {Dimmelmeier}, {Marek}, {Janka}, {Zink}, {Hawke}  \& {Schnetter}}{{Ott} et~al.}{2007}]{ott:07cqg}
{Ott} C.~D.,  {Dimmelmeier} H.,  {Marek} A.,  {Janka} H.~T.,  {Zink} B.,  {Hawke} I.,   {Schnetter} E.,  2007, \mn@doi [Classical and Quantum Gravity] {10.1088/0264-9381/24/12/S10}, \href {https://ui.adsabs.harvard.edu/abs/2007CQGra..24S.139O} {24, S139}

\bibitem[\protect\citeauthoryear{{Ott} et~al.,}{{Ott} et~al.}{2012}]{ott:12}
{Ott} C.~D.,  et~al., 2012, \mn@doi [\prd] {10.1103/PhysRevD.86.024026}, \href {https://ui.adsabs.harvard.edu/abs/2012PhRvD..86b4026O} {86, 024026}

\bibitem[\protect\citeauthoryear{Pacilio, Maselli, Fasano  \& Pani}{Pacilio et~al.}{2022}]{pacilio2022ranking}
Pacilio C.,  Maselli A.,  Fasano M.,   Pani P.,  2022, Physical Review Letters, 128, 101101

\bibitem[\protect\citeauthoryear{{Pajkos}, {Warren}, {Couch}, {O'Connor}  \& {Pan}}{{Pajkos} et~al.}{2021}]{pajkos21}
{Pajkos} M.~A.,  {Warren} M.~L.,  {Couch} S.~M.,  {O'Connor} E.~P.,   {Pan} K.-C.,  2021, \mn@doi [\apj] {10.3847/1538-4357/abfb65}, \href {https://ui.adsabs.harvard.edu/abs/2021ApJ...914...80P} {914, 80}

\bibitem[\protect\citeauthoryear{Pastor-Marcos, Cerd{\'a}-Dur{\'a}n, Walker, Torres-Forn{\'e}, Abdikamalov, Richers  \& Font}{Pastor-Marcos et~al.}{2023}]{pastor2023bayesian}
Pastor-Marcos C.,  Cerd{\'a}-Dur{\'a}n P.,  Walker D.,  Torres-Forn{\'e} A.,  Abdikamalov E.,  Richers S.,   Font J.~A.,  2023, arXiv preprint arXiv:2308.03456

\bibitem[\protect\citeauthoryear{Pedregosa et~al.,}{Pedregosa et~al.}{2018}]{pedregosa2018scikitlearn}
Pedregosa F.,  et~al., 2018, Scikit-learn: Machine Learning in Python (\mn@eprint {arXiv} {1201.0490})

\bibitem[\protect\citeauthoryear{Pinkus}{Pinkus}{1999}]{pinkus1999approximation}
Pinkus A.,  1999, Acta Numerica, 8, 143

\bibitem[\protect\citeauthoryear{{Popov} \& {Turolla}}{{Popov} \& {Turolla}}{2012}]{popov:12}
{Popov} S.~B.,  {Turolla} R.,  2012, \mn@doi [\apss] {10.1007/s10509-012-1100-z}, \href {https://ui.adsabs.harvard.edu/abs/2012Ap&SS.341..457P} {341, 457}

\bibitem[\protect\citeauthoryear{{Powell} \& {M{\"u}ller}}{{Powell} \& {M{\"u}ller}}{2020}]{powell:20}
{Powell} J.,  {M{\"u}ller} B.,  2020, \mn@doi [\mnras] {10.1093/mnras/staa1048}, \href {https://ui.adsabs.harvard.edu/abs/2020MNRAS.494.4665P} {494, 4665}

\bibitem[\protect\citeauthoryear{{Powell} \& {M{\"u}ller}}{{Powell} \& {M{\"u}ller}}{2022}]{Powell22}
{Powell} J.,  {M{\"u}ller} B.,  2022, \mn@doi [\prd] {10.1103/PhysRevD.105.063018}, \href {https://ui.adsabs.harvard.edu/abs/2022PhRvD.105f3018P} {105, 063018}

\bibitem[\protect\citeauthoryear{{Powell}, {Gossan}, {Logue}  \& {Heng}}{{Powell} et~al.}{2016}]{Powell16}
{Powell} J.,  {Gossan} S.~E.,  {Logue} J.,   {Heng} I.~S.,  2016, \mn@doi [\prd] {10.1103/PhysRevD.94.123012}, \href {https://ui.adsabs.harvard.edu/abs/2016PhRvD..94l3012P} {94, 123012}

\bibitem[\protect\citeauthoryear{Puecher, Dietrich, Tsang, Kalaghatgi, Roy, Setyawati  \& Van Den~Broeck}{Puecher et~al.}{2023}]{puecher2023unraveling}
Puecher A.,  Dietrich T.,  Tsang K.~W.,  Kalaghatgi C.,  Roy S.,  Setyawati Y.,   Van Den~Broeck C.,  2023, Physical Review D, 107, 124009

\bibitem[\protect\citeauthoryear{{Radice}, {Ott}, {Abdikamalov}, {Couch}, {Haas}  \& {Schnetter}}{{Radice} et~al.}{2016}]{radice:16a}
{Radice} D.,  {Ott} C.~D.,  {Abdikamalov} E.,  {Couch} S.~M.,  {Haas} R.,   {Schnetter} E.,  2016, \mn@doi [\apj] {10.3847/0004-637X/820/1/76}, \href {http://adsabs.harvard.edu/abs/2016ApJ...820...76R} {820, 76}

\bibitem[\protect\citeauthoryear{Radice, Perego, Zappa  \& Bernuzzi}{Radice et~al.}{2018}]{radice2018gw170817}
Radice D.,  Perego A.,  Zappa F.,   Bernuzzi S.,  2018, The Astrophysical Journal Letters, 852, L29

\bibitem[\protect\citeauthoryear{{Radice}, {Morozova}, {Burrows}, {Vartanyan}  \& {Nagakura}}{{Radice} et~al.}{2019}]{radice:19gw}
{Radice} D.,  {Morozova} V.,  {Burrows} A.,  {Vartanyan} D.,   {Nagakura} H.,  2019, \mn@doi [\apjl] {10.3847/2041-8213/ab191a}, \href {https://ui.adsabs.harvard.edu/abs/2019ApJ...876L...9R} {876, L9}

\bibitem[\protect\citeauthoryear{{Richers}, {Ott}, {Abdikamalov}, {O'Connor}  \& {Sullivan}}{{Richers} et~al.}{2017}]{richers:17}
{Richers} S.,  {Ott} C.~D.,  {Abdikamalov} E.,  {O'Connor} E.,   {Sullivan} C.,  2017, \mn@doi [\prd] {10.1103/PhysRevD.95.063019}, \href {https://ui.adsabs.harvard.edu/abs/2017PhRvD..95f3019R} {95, 063019}

\bibitem[\protect\citeauthoryear{{Sagert}, {Fischer}, {Hempel}, {Pagliara}, {Schaffner-Bielich}, {Mezzacappa}, {Thielemann}  \& {Liebend{\"o}rfer}}{{Sagert} et~al.}{2009}]{sagert:09}
{Sagert} I.,  {Fischer} T.,  {Hempel} M.,  {Pagliara} G.,  {Schaffner-Bielich} J.,  {Mezzacappa} A.,  {Thielemann} F.,   {Liebend{\"o}rfer} M.,  2009, \prl, \href {http://adsabs.harvard.edu/abs/2009PhRvL.102h1101S} {102, 081101}

\bibitem[\protect\citeauthoryear{{Saiz-P{\'e}rez}, {Torres-Forn{\'e}}  \& {Font}}{{Saiz-P{\'e}rez} et~al.}{2022}]{Saiz-Perez22}
{Saiz-P{\'e}rez} A.,  {Torres-Forn{\'e}} A.,   {Font} J.~A.,  2022, \mn@doi [\mnras] {10.1093/mnras/stac698}, \href {https://ui.adsabs.harvard.edu/abs/2022MNRAS.512.3815S} {512, 3815}

\bibitem[\protect\citeauthoryear{{Scheck}, {Janka}, {Foglizzo}  \& {Kifonidis}}{{Scheck} et~al.}{2008}]{scheck:08}
{Scheck} L.,  {Janka} H.-T.,  {Foglizzo} T.,   {Kifonidis} K.,  2008, \aap, 477, 931

\bibitem[\protect\citeauthoryear{{Shen}, {Horowitz}  \& {O'Connor}}{{Shen} et~al.}{2011}]{gshen:11b}
{Shen} G.,  {Horowitz} C.~J.,   {O'Connor} E.,  2011, \mn@doi [Phys.\ Rev.\ C] {10.1103/PhysRevC.83.065808}, \href {http://adsabs.harvard.edu/abs/2011PhRvC..83f5808S} {83, 065808}

\bibitem[\protect\citeauthoryear{Simonyan \& Zisserman}{Simonyan \& Zisserman}{2014}]{simonyan2014very}
Simonyan K.,  Zisserman A.,  2014, arXiv preprint arXiv:1409.1556

\bibitem[\protect\citeauthoryear{Smith \& Gossett}{Smith \& Gossett}{1984}]{SmithGossett1984}
Smith J.,  Gossett P.,  1984, in ICASSP '84. IEEE International Conference on Acoustics, Speech, and Signal Processing. IEEE, pp 112--115

\bibitem[\protect\citeauthoryear{S{\o}nderby, Raiko, Maal{\o}e  \& S{\o}nderby}{S{\o}nderby et~al.}{2016}]{sonderby2016categorical}
S{\o}nderby C.~K.,  Raiko T.,  Maal{\o}e L.,   S{\o}nderby S.~K.,  2016, in Proceedings of the 33rd International Conference on International Conference on Machine Learning - Volume 48. JMLR.org

\bibitem[\protect\citeauthoryear{{Sotani}, {Takiwaki}  \& {Togashi}}{{Sotani} et~al.}{2021}]{Sotani21}
{Sotani} H.,  {Takiwaki} T.,   {Togashi} H.,  2021, \mn@doi [\prd] {10.1103/PhysRevD.104.123009}, \href {https://ui.adsabs.harvard.edu/abs/2021PhRvD.104l3009S} {104, 123009}

\bibitem[\protect\citeauthoryear{Srivastava, Ballmer, Brown, Afle, Burrows, Radice  \& Vartanyan}{Srivastava et~al.}{2019}]{Srivastava:2019fcb}
Srivastava V.,  Ballmer S.,  Brown D.~A.,  Afle C.,  Burrows A.,  Radice D.,   Vartanyan D.,  2019, \mn@doi [Phys. Rev. D] {10.1103/PhysRevD.100.043026}, 100, 043026

\bibitem[\protect\citeauthoryear{{Steiner}, {Hempel}  \& {Fischer}}{{Steiner} et~al.}{2013}]{steiner:13b}
{Steiner} A.~W.,  {Hempel} M.,   {Fischer} T.,  2013, \mn@doi [\apj] {10.1088/0004-637X/774/1/17}, \href {http://adsabs.harvard.edu/abs/2013ApJ...774...17S} {774, 17}

\bibitem[\protect\citeauthoryear{Sullivan, O'Connor, Zegers, Grubb  \& Austin}{Sullivan et~al.}{2016}]{Sullivan16}
Sullivan C.,  O'Connor E.,  Zegers R. G.~T.,  Grubb T.,   Austin S.~M.,  2016, \mn@doi [Astrophys. J.] {10.3847/0004-637X/816/1/44}, 816, 44

\bibitem[\protect\citeauthoryear{{Szczepa{\'n}czyk} et~al.,}{{Szczepa{\'n}czyk} et~al.}{2021}]{Szczepanczyk21}
{Szczepa{\'n}czyk} M.~J.,  et~al., 2021, \mn@doi [\prd] {10.1103/PhysRevD.104.102002}, \href {https://ui.adsabs.harvard.edu/abs/2021PhRvD.104j2002S} {104, 102002}

\bibitem[\protect\citeauthoryear{Szczepa{\'n}czyk et~al.,}{Szczepa{\'n}czyk et~al.}{2023}]{szczepanczyk2023optically}
Szczepa{\'n}czyk M.~J.,  et~al., 2023, arXiv preprint arXiv:2305.16146

\bibitem[\protect\citeauthoryear{Szegedy et~al.,}{Szegedy et~al.}{2015}]{szegedy2015going}
Szegedy C.,  et~al., 2015, in Proceedings of the IEEE conference on computer vision and pattern recognition. pp~1--9

\bibitem[\protect\citeauthoryear{{Torres-Forn{\'e}}, {Cerd{\'a}-Dur{\'a}n}, {Passamonti}, {Obergaulinger}  \& {Font}}{{Torres-Forn{\'e}} et~al.}{2019}]{Torres-Forne19}
{Torres-Forn{\'e}} A.,  {Cerd{\'a}-Dur{\'a}n} P.,  {Passamonti} A.,  {Obergaulinger} M.,   {Font} J.~A.,  2019, \mn@doi [\mnras] {10.1093/mnras/sty2854}, \href {https://ui.adsabs.harvard.edu/abs/2019MNRAS.482.3967T} {482, 3967}

\bibitem[\protect\citeauthoryear{{Varma}, {M{\"u}ller}  \& {Schneider}}{{Varma} et~al.}{2023}]{varma23}
{Varma} V.,  {M{\"u}ller} B.,   {Schneider} F. R.~N.,  2023, \mn@doi [\mnras] {10.1093/mnras/stac3247}, \href {https://ui.adsabs.harvard.edu/abs/2023MNRAS.518.3622V} {518, 3622}

\bibitem[\protect\citeauthoryear{{Vartanyan}, {Burrows}, {Wang}, {Coleman}  \& {White}}{{Vartanyan} et~al.}{2023}]{Vartanyan23}
{Vartanyan} D.,  {Burrows} A.,  {Wang} T.,  {Coleman} M. S.~B.,   {White} C.~J.,  2023, \mn@doi [\prd] {10.1103/PhysRevD.107.103015}, \href {https://ui.adsabs.harvard.edu/abs/2023PhRvD.107j3015V} {107, 103015}

\bibitem[\protect\citeauthoryear{Wang, Yan  \& Oates}{Wang et~al.}{2017}]{wang2017time}
Wang Z.,  Yan W.,   Oates T.,  2017, in 2017 International Joint Conference on Neural Networks (IJCNN). pp 1578--1585

\bibitem[\protect\citeauthoryear{{Warren}, {Couch}, {O'Connor}  \& {Morozova}}{{Warren} et~al.}{2020}]{Warren20}
{Warren} M.~L.,  {Couch} S.~M.,  {O'Connor} E.~P.,   {Morozova} V.,  2020, \mn@doi [\apj] {10.3847/1538-4357/ab97b7}, \href {https://ui.adsabs.harvard.edu/abs/2020ApJ...898..139W} {898, 139}

\bibitem[\protect\citeauthoryear{{Winteler}, {K{\"a}ppeli}, {Perego}, {Arcones}, {Vasset}, {Nishimura}, {Liebend{\"o}rfer}  \& {Thielemann}}{{Winteler} et~al.}{2012}]{winteler12}
{Winteler} C.,  {K{\"a}ppeli} R.,  {Perego} A.,  {Arcones} A.,  {Vasset} N.,  {Nishimura} N.,  {Liebend{\"o}rfer} M.,   {Thielemann} F.~K.,  2012, \mn@doi [\apjl] {10.1088/2041-8205/750/1/L22}, \href {https://ui.adsabs.harvard.edu/abs/2012ApJ...750L..22W} {750, L22}

\bibitem[\protect\citeauthoryear{{Wolfe}, {Fr{\"o}hlich}, {Miller}, {Torres-Forn{\'e}}  \& {Cerd{\'a}-Dur{\'a}n}}{{Wolfe} et~al.}{2023}]{Wolfe23}
{Wolfe} N.~E.,  {Fr{\"o}hlich} C.,  {Miller} J.~M.,  {Torres-Forn{\'e}} A.,   {Cerd{\'a}-Dur{\'a}n} P.,  2023, \mn@doi [\apj] {10.3847/1538-4357/ace693}, \href {https://ui.adsabs.harvard.edu/abs/2023ApJ...954..161W} {954, 161}

\bibitem[\protect\citeauthoryear{{Woosley} \& {Bloom}}{{Woosley} \& {Bloom}}{2006}]{woosley_bloom:06}
{Woosley} S.~E.,  {Bloom} J.~S.,  2006, \mn@doi [\araa] {10.1146/annurev.astro.43.072103.150558}, \href {https://ui.adsabs.harvard.edu/abs/2006ARA&A..44..507W} {44, 507}

\bibitem[\protect\citeauthoryear{{Woosley} \& {Heger}}{{Woosley} \& {Heger}}{2006}]{woosley:06}
{Woosley} S.~E.,  {Heger} A.,  2006, \apj, 637, 914

\bibitem[\protect\citeauthoryear{{Woosley} \& {Heger}}{{Woosley} \& {Heger}}{2007}]{woosley:07}
{Woosley} S.~E.,  {Heger} A.,  2007, \physrep, \href {http://adsabs.harvard.edu/abs/2007PhR...442..269W} {442, 269}

\bibitem[\protect\citeauthoryear{{Woosley}, {Heger}  \& {Weaver}}{{Woosley} et~al.}{2002}]{whw:02}
{Woosley} S.~E.,  {Heger} A.,   {Weaver} T.~A.,  2002, Rev.\ Mod.\ Phys., 74, 1015

\bibitem[\protect\citeauthoryear{{Yahil}}{{Yahil}}{1983}]{yahil:83}
{Yahil} A.,  1983, \apj, 265, 1047

\bibitem[\protect\citeauthoryear{{Yoon}, {Langer}  \& {Norman}}{{Yoon} et~al.}{2006}]{yoon:06}
{Yoon} S.-C.,  {Langer} N.,   {Norman} C.,  2006, \mn@doi [\aap] {10.1051/0004-6361:20065912}, \href {http://adsabs.harvard.edu/abs/2006A%26A...460..199Y} {460, 199}

\bibitem[\protect\citeauthoryear{{Yuan}, {Fan}, {Lv}, {Sun}  \& {Lin}}{{Yuan} et~al.}{2023}]{Yuan23reconstruction}
{Yuan} Y.,  {Fan} X.-L.,  {Lv} H.-J.,  {Sun} Y.-Y.,   {Lin} K.,  2023, \mn@doi [arXiv e-prints] {10.48550/arXiv.2309.06011}, \href {https://ui.adsabs.harvard.edu/abs/2023arXiv230906011Y} {p. arXiv:2309.06011}

\bibitem[\protect\citeauthoryear{{Zha}, {O'Connor}  \& {da Silva Schneider}}{{Zha} et~al.}{2021}]{zha:21}
{Zha} S.,  {O'Connor} E.~P.,   {da Silva Schneider} A.,  2021, \mn@doi [\apj] {10.3847/1538-4357/abec4c}, \href {https://ui.adsabs.harvard.edu/abs/2021ApJ...911...74Z} {911, 74}

\bibitem[\protect\citeauthoryear{{da Silva Schneider}, {O'Connor}, {Granqvist}, {Betranhandy}  \& {Couch}}{{da Silva Schneider} et~al.}{2020}]{daSilvaSchneider20}
{da Silva Schneider} A.,  {O'Connor} E.,  {Granqvist} E.,  {Betranhandy} A.,   {Couch} S.~M.,  2020, \mn@doi [\apj] {10.3847/1538-4357/ab8308}, \href {https://ui.adsabs.harvard.edu/abs/2020ApJ...894....4D} {894, 4}

\bibitem[\protect\citeauthoryear{{de Mink}, {Langer}, {Izzard}, {Sana}  \& {de Koter}}{{de Mink} et~al.}{2013}]{demink:13}
{de Mink} S.~E.,  {Langer} N.,  {Izzard} R.~G.,  {Sana} H.,   {de Koter} A.,  2013, \apj, 764, 166

\makeatother
\end{thebibliography}

%%%%%%%%%%%%%%%%%%%%%%%%%%%%%%%%%%%%%%%%%%%%%%%%%%

%%%%%%%%%%%%%%%%% APPENDICES %%%%%%%%%%%%%%%%%%%%%

\appendix

\section{Fourier Space Analysis}
\label{sec:fourier}

In this section, we provide a classification analysis in the Fourier space. We use the same neural network pipeline described in Section \ref{sec:method}. Our analysis is based on computing the amplitude spectral density (ASD) from a power spectral density (PSD) using a periodogram approach \citep{abbot16PRL, aasi2015advanced, allen1998data}. It is the squared magnitude of the Fourier transform of a sequence, normalized by the length of the sequence. A periodogram can be viewed as the squared modulus of the Fourier coefficients, representing the power of each frequency component in the data \citep{Bloomfield2004}. Given a sequence of observations $x_1, x_2, \dots, x_N$, the periodogram $I(f)$ is defined as
\[
I(f) = \frac{1}{N} \left| \sum_{n=1}^{N} x_n e^{-i 2\pi f n} \right|^2,
\]
where $f$ is the frequency and $N$ is the number of observations. 
The ASD is given as
\[
ASD = \sqrt{I(f)}
\]

We train our CNN model using the ASD data for groups 1, 2, 3, and 3b (see Section~\ref{sec:method} for the description of these groups). The corresponding accuracies are \gfirstFTscore, \gsecFTscore, \gthirdFTscore, and \gthirdbFTscore. In agreement with the results obtained using the time-series data, group 3 exhibits the highest accuracy, while groups 2, 3b, and 1 have accuracy values in descending order, as shown in Fig.~\ref{fig:f1}. Despite this correspondence, the Fourier space accuracies are on average $\sim 3 $ percent lower than the corresponding accuracies in the time-space. A drop in accuracy was also observed by \cite{Edwards21}. Below, we outline a few factors that could contribute to this behavior. However, note that this list may not be exhaustive.

First, the patterns and dynamic range of features in the time and frequency domains could be different from each other. The ASD can compress or expand certain features, making them less distinguishable for the neural network \citep[e.g.,][]{fawaz2019deep}. Moreover, transforming the data from the time domain to the frequency domain could result in a loss of some signal features, especially those that are temporally localized \citep[e.g.,][]{george2018deep}.

Second, the CNN parameters that we use may be more suited to time-series data than frequency-domain data. Additional adjustments may be needed when working in Fourier space. Also, it is possible that the training dynamics, such as the learning rate, batch size, or regularization, which are optimized for the time-series data, may not be as suitable for the ASD representation \citep[e.g.,][]{wang2017time}. A detailed analysis of these aspects will be the subject of a future work.

\section{Analysis of Data Sampling Rates}
\label{sec:sampling}

\begin{figure}
    \centering
    \includegraphics[scale=0.49]{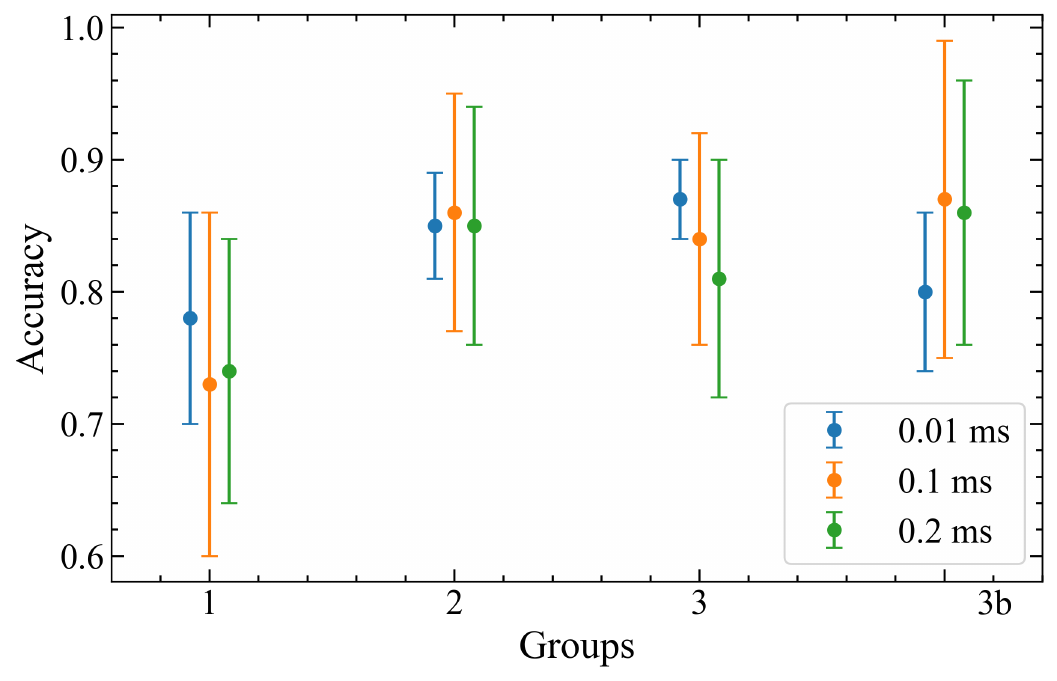}
    \caption{EOS classification accuracy for groups 1, 2, 3, and 3b using different data sampling rates. The blue dots represent a sampling rate of $0.01$ ms, orange dots indicate a sampling rate of $0.1$ ms, and green dots denote a sampling rate of $0.2$ ms.}
    \label{fig:sampling}
\end{figure}

In this section we present  a supplementary check on the effect of classification accuracy as a function of the number of sampled points. For this purpose, we choose three sampling rates: $0.2$, $0.1$, and $0.01$ ms for each group. We then compute the classification accuracy for each of the four groups, which are shown in Fig.~\ref{fig:sampling}. We can see that the accuracy scores are in line with each other within the error bars. 

\begin{figure}
    \centering
    \includegraphics[scale=0.44]{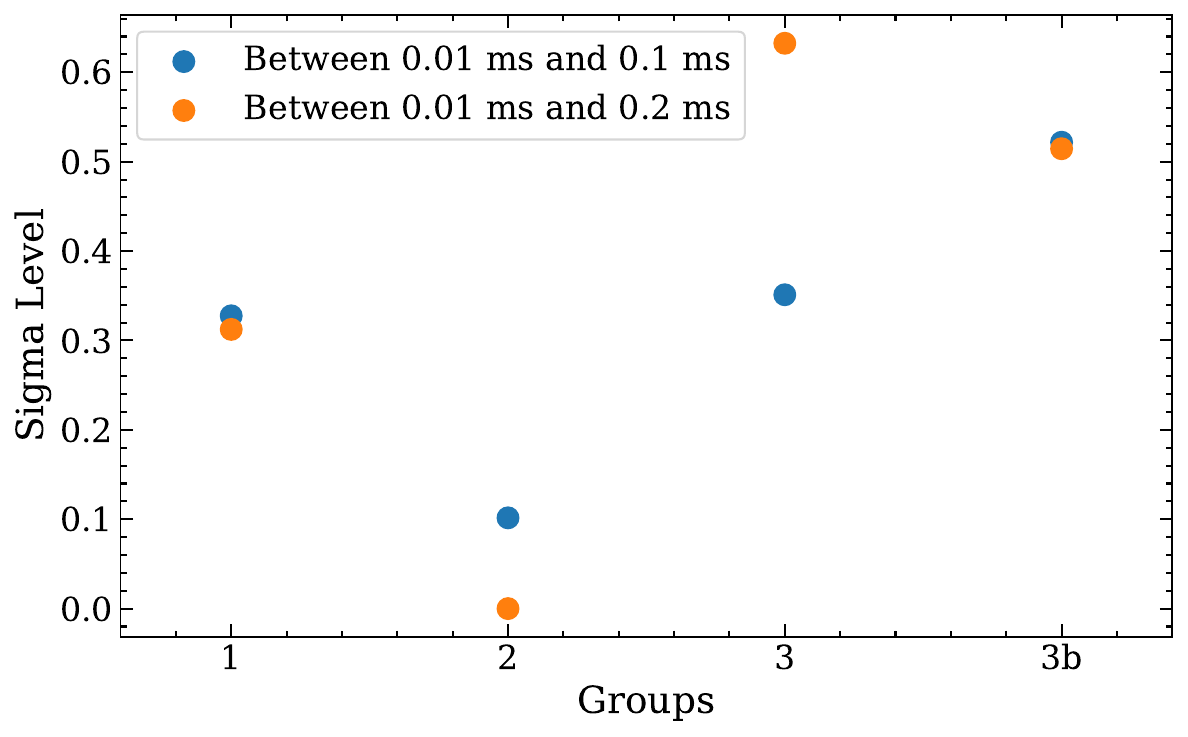}
    \caption{Sigma deviation levels for accuracy values between the $0.01$ ms sampling dataset and the $0.2$ and $0.1$ ms sampling datasets for each group. All sigma deviation values are less than $1$, which suggests that there is no statistically significant dependence on EOS classification results as a function of the sampling frequency.}
    \label{fig:sigma_significance}
\end{figure}

To further support this finding, we compute the sigma levels of deviation, given the mean and standard deviation of accuracy values for each groups. The sigma level of deviation is computed for two observations, $x$ and $y$, as follows
\begin{equation}
\Delta = \frac{|x - y|}{\sqrt{\sigma_x^2 + \sigma_y^2}}.
\end{equation}
As shown in Fig.~\ref{fig:sigma_significance}, all the sigma levels are smaller than $1$, signifying that the sampling frequency of the data does not significantly impact the EOS classification accuracy. Based on this finding, throughout the paper we perform our analysis with the $0.01$ ms sampling dataset, as it represents the largest dataset, which we can execute within reasonable computational resources.

%%%%%%%%%%%%%%%%%%%%%%%%%%%%%%%%%%%%%%%%%%%%%%%%%%

\bsp	% typesetting comment
\label{lastpage}
\end{document}